\documentclass{aa}  

\usepackage{graphicx}
\usepackage{txfonts}
\usepackage{natbib}
\bibpunct{(}{)}{;}{a}{}{,} 
\pdfoutput=0
\begin{document}

\title{
\textit{Herschel}\thanks{The \textit{Herschel} data described in this paper
have been obtained in the open time project {\tt OT1\_tpreibis\_1} (PI: T.~Preibisch). \textit{Herschel}
is an ESA space observatory with science instruments provided by European-led Principal Investigator
consortia and with important participation from NASA.}
 far-infrared observations of the Carina Nebula Complex}

\subtitle{ {\rm III:} Detailed cloud structure and feedback effects}

   \author{V.~Roccatagliata\inst{1},
	T.~Preibisch\inst{1},
	T.~Ratzka\inst{1},
	 \and
         B.~Gaczkowski\inst{1}
          }

\institute{Universit{\"a}ts-Sternwarte M{\"u}nchen, Ludwig-Maximilians-Universit{\"a}t,
Scheinerstr.~1, 81679 M{\"u}nchen, Germany \email{vrocca@usm.uni-muenchen.de} }

   \date{Received January 11, 2013; Accepted March 19, 2013}

  \abstract
   {
The star formation process in large clusters/associations can be strongly influenced by the 
feedback from high mass stars. Whether the resulting net effect of the feedback is
predominantly negative (cloud dispersal) or positive (triggering of star formation
due to cloud compression) is still an open question.
   }
   {
The Carina Nebula complex (CNC) represents one of the most massive star-forming regions in our Galaxy.
We use our \textit{Herschel} far-infrared observations to 
study the properties of the clouds 
 over the entire area of the CNC (with a diameter of  $\approx 3.2\degr$,
which corresponds to $\approx 125$~pc at the distance of 2.3~kpc).
The good angular resolution ($10'' - 36''$) of the \textit{Herschel} maps
corresponds to physical scales of 0.1 -- 0.4~pc, and
allows us to analyze the small-scale (i.e.~clump-size) structures of the clouds.
   }
   {
The full extent of the CNC was mapped with PACS and SPIRE in the 70, 160, 250, 350, and $500\,\mu$m bands.
 We determine temperatures and column densities at each point in this maps by modeling the observed
 far-infrared spectral energy distributions. We also derive a map showing the strength of the
UV radiation field. We investigate the relation between the cloud properties and the spatial distribution
of the high-mass stars, and compute total cloud masses for different density thresholds.}
   {Our \textit{Herschel} maps resolve, for the first time, the small-scale structure of the 
dense clouds over the
entire spatial extent of the CNC. Several particularly interesting
   regions, including the prominent pillars south of $\eta$~Car, 
   are analyzed in detail.  
  We compare the cloud masses derived from the \textit{Herschel} data to previous mass estimates
  based on sub-mm and molecular line data.
Our maps also reveal a peculiar ``wave''-like pattern 
in the northern part of the Carina Nebula.
Finally, we characterize two prominent cloud
   complexes at the periphery of our \textit{Herschel} maps, which are probably molecular clouds in the
   Galactic background. 
   }
   {
We find that the
density and temperature structure of the clouds in most parts of the CNC is dominated 
by the strong feedback from the numerous massive stars, rather  
 than random turbulence.
Comparing the cloud mass and the star formation rate derived for the CNC to other Galactic star forming 
regions suggests that the CNC is forming stars in an particularly efficient way.
We suggest  this to be a consequence of triggered star formation by radiative cloud compression.
   }


   \titlerunning{Cloud structure and feedback in the Carina Nebula complex} 
   \authorrunning{Roccatagliata et al.}

   \maketitle
   %

\section{Introduction}

The Carina Nebula complex (CNC hereafter) \citep[e.g. ][for an overview of the region]{SmithBrooks2008}
is one of the richest and largest high-mass star forming regions in our Galaxy. At a moderate distance of 
2.3~kpc, it contains at least 65 O-type stars \citep[][]{Smith2006} and four Wolf-Rayet stars 
\citep[see e.g.][]{SmithConti2008}. \\
The CNC extends over at least $\sim 80$~pc, which corresponds to 
more than 2 degrees on the sky, demonstrating that wide-field surveys are necessary in order to obtain a 
comprehensive information on the full star forming complex.
Several wide-field surveys of the CNC have recently been carried out at different wavelengths.
The combination of a large {\it Chandra} X-ray survey \citep[see][]{Townsleyetal2011}
with a deep near-infrared survey \citet{Preibischetal2011d, Preibischetal2011b}
and \textit{Spitzer} mid-infrared observations \citep[][]{Smithetal2010, Povichetal2011}
provided comprehensive information about the young stellar populations.
Our sub-millimeter survey of the CNC with LABOCA at the APEX telescope
revealed the structure of the cold and dense clouds in the complex in detail \citep[][]{Preibischetal2011a}.

The CNC is an interesting region in which to study the feedback effects of the numerous very massive
and luminous stars on the surrounding clouds. The strong ionizing radiation and the powerful winds of 
the high-mass stars affect the clouds in very different ways: on the one hand, the process of
 photoevaporation at strongly irradiated cloud surfaces can disperse even rather massive clouds on 
relatively short timescales; transforming dense molecular clouds into warm/hot low-density atomic gas
will strongly limit the potential for further star formation. 
On the other hand, the compression of clouds by irradiation and by expanding bubbles
driven by evolving HII regions \citep[e.g.][]{Deharvengetal2005} or stellar winds can produce gravitationally 
unstable density peaks and lead to triggered star formation \citep[e.g.][]{Smithetal2010}. 
The detailed balance of these two 
opposing processes determines the evolution of the complex and decides, how much of the original 
cloud mass is actually transformed into stars and what fraction is dispersed by the feedback.
In order to study this interaction between the stars and the surrounding clouds, a comprehensive
characterization of the cloud structure and temperature is a fundamental 
requirement.

The ESA \textit{Herschel} Space Observatory \citep{Pilbrattetal2010} is ideally suited to map
the far-infrared emission of the warm and cool/cold molecular clouds  and is currently observing 
many galactic (and extragalactic) star forming regions.
We have used the \textit{Herschel} Observatory to perform a wide-field ($\ga 5$~square-degrees) survey
of the CNC (P.I.: Th.~Preibisch) that covers the entire spatial extent of the clouds.
Far-infrared photometric maps were obtained with PACS 
\citep{Poglitschetal2010} at 70 and 160 $\mu$m, and with SPIRE \citep{Griffinetal2010} at 250, 350, and $500~\mu$m.
A first analysis of these \textit{Herschel} observations, focusing on the large scale structures and
global properties, has been presented  in \citet{Preibischetal2012} (Paper~I in the following text). 
A detailed study of the 
young stellar and protostellar population detected as point-sources in the \textit{Herschel} maps is
described in \citet{Gaczkowskietal2013}. The properties of the clouds around the
prominent HII region Gum~31, located just north-west of the central Carina Nebula,
as derived from these \textit{Herschel} and other observations are studied in \citet{Ohlendorfetal2013}.
In this paper, we present a detailed analysis of the temperature and column density of the small-scale 
structures of the entire CNC.

This paper is organized as follows: in Section~\ref{obs} we briefly summarize the {\it Herschel} PACS and SPIRE 
observations, data reduction and the final temperature and column density maps of the CNC. 
In Section~\ref{an} we compute the temperature, column density and UV-flux maps of the clouds. 
In Section~\ref{clouds} we present a detailed analysis of the cloud structures resolved by {\it Herschel} 
in the CNC., e.g. the southern pillars.  
We discuss, in Section~\ref{discussion}, the comparison between dense and diffuse gas, 
the relation between dust and gas mass estimates, the strength of the radiative feedback and  
the CNC as a link between local and extragalactic star formation. 
A summary of our results and conclusions is given in Section~\ref{conclusions}.

\section{The Herschel PACS and SPIRE maps of the Carina Nebula Complex}
\label{obs}

\subsection{Observations and Data Reduction}
The CNC has been observed on December 26, 2010 using the parallel fast scan mode at $60''$/s, obtaining 
simultaneously $70\,\mu$m and $160\,\mu$m images with PACS, and 250, 350, and 500 $\mu$m maps with SPIRE. 

The scan maps cover an area of 2.8$^\circ\times2.8^\circ$, which corresponds to about 110\,pc\,$\times$\,110\,pc 
at the distance of the CNC.
The data reduction has been carried out combining the HIPE \citep[v. 7.0;][]{Ott2010}
with the {\tt SCANAMORPHOS} package \citep[v. 13.0;][]{Roussel2012}.
All the details of the reduction can be found in Paper I. 
We here only emphasize that we used the option {\tt galactic} in {\tt SCANAMORPHOS} which preserves 
the brightness gradient over the field. 
The final angular resolution of the PACS/\textit{Herschel}\, maps was $\approx 12'' - 16''$,  
while for the SPIRE/\textit{Herschel}\, maps $\approx 20'' - 36''$.

\subsection{Calibration}
\label{res}
The observations and mosaics of the CNC have been previously presented in Paper I. 
In this work an additional effort to calibrate the final mosaic has been 
done by discussing the effect of the {\it large scale flux loss} and taking into account  the 
{\it color correction}, which might influence our further analysis.
In the first case, 
it is important to recall that {\it Herschel} is a warm instrument, not designed to obtain absolute fluxes.  
The warm emission from the telescope's mirror is removed using a high-pass filtering 
during the data reduction. However, this step might remove some large scale emission in the final 
mosaics. 
The flux loss can only be estimated by extrapolating the {\it Herschel} fluxes to the wavelengths 
observed by a cold telescope, such as e.g. {\it Planck}. 

Another possibility is to obtain the contributions from background and foreground material estimating the dust in 
foreground/background gas that is traced by H\,I, for the atomic emission, and CO, for the 
molecular emission. This approach has been presented by \citet{Rivera-Ingrahametal2013}. 

Since the {\it Planck} data are not yet publicly available and we do not have H\,I and CO  
spectral data of the CNC, we compared our {\it Herschel} 70~$\mu$m map 
to the IRAS 60~$\mu$m map. This approach was already used in Paper I and the details can be found 
there.

We also use another approach, which consists in removing the local background at each wavelength, 
by assuming an average value over the field \citep[see e.g. ][]{Stutzetal2010}. 
In this way we only take into account the {\it Herschel} emission above this local 
background. This was computed in a region of the PACS and SPIRE maps outside the CNC where there 
was no cloud emission. We find that the local background is 0.0005 Jy/arcsec$^2$.

We checked whether the subtraction of the local background estimated was affecting our result on 
the temperature structure.

Both approaches show that the {\it large scale flux loss} can introduce an error negligible compared to 
a conservative calibration error of 20\% of the flux adopted in our work.

The second calibration step taken into account is the {\it color correction}. 
This  correction can affect the Herschel photometry at the shortest wavelengths in the 
case of low temperatures\footnote{PACS manual: {\it ``PACS Photometer Passbands and Colour Correction Factors for Various Source SEDs''}}
To correct the PACS fluxes we used the first temperature map obtained by fitting the SED on the 
uncorrected {\it Herschel} fluxes ($F_{unc}$). For a given 
temperature, we applied the color corrections $cc_{pacs70}$ and $cc_{pacs160}$ listed in 
the PACS manual$^1$; 
 the corrected flux $F_{corr}$ is computed as $F_{corr}=F_{unc}/cc_{pacs}$.

In a second step the fit of the SED has been repeated using  the color-corrected fluxes.

Between the two iterations the temperature map differs of about 0.1\,K in the coldest part of the CNC 
(T$\sim$16~K). For warmer parts the correction was less than 0.1\,K.

We checked the effect of the color-correction on the SPIRE photometry. 
The SPIRE color corrections are obtained checking the slope $\alpha$ of the 
spectral energy distribution (SED hereafter) between  250 and 500~$\mu$m. 
From an average spectral index value of about 2.4 over the entire {\it Herschel} mosaic, 
the color correction would have ranged between 1.02 and 0.97 \footnote{SPIRE color corrections from the SPIRE 
Photometry Cookbook; Bendo et al. (2011)}. This correction is hence negligible.


\section{Analysis}
\label{an}
\subsection{Temperature and column density of the clouds}
In this section we describe in detail the procedure to obtain a precise temperature and column density map 
of the CNC.
The most reliable and powerful approach to derive the dust temperature and density maps is fitting, pixel by pixel,  
the SED to the fluxes in the 5 {\it Herschel} bands. 
The final {\it Herschel} mosaics at 70, 160, 250, 350, and 500 $\mu$m (derived as previously described) 
have been convolved to the angular resolution of the 500~$\mu$m image, using the procedure presented in 
\citet{Stutzetal2010} and the kernels from \citet{Anianoetal2011}.

Since the nebula is optically thin at all the {\it Herschel} wavelengths \citep[][]{Preibischetal2011b}, the emission of the 
nebula is $B_\nu(T)\cdot \tau$, where $B_\nu$ is the Planck function at a temperature $T$ and 
$\tau$ is the optical depth, which is proportional 
to the dust mass absorption coefficient, $k_\nu$. 
From the models presented in \citet{OssenkopfHenning1994}, we choose a dust model 
with a highest $\beta$ value (i.e. 1.9) which is representative to describe a dense molecular cloud. 
This model has a standard MRN-distribution for diffuse interstellar medium \citep[][]{DraineLee1984} 
composed of grains without ice mantels. 
We used the tabulated values of $k_\nu$ [cm$^2$/g] 
for such a model and interpolated at the {\it Herschel} wavelengths. 
In Table~\ref{kappa} we show the values of $k_\nu$ that we obtained.
\begin{table}
\caption{Values of the dust mass absorption coefficient, $k_\nu$ interpolated at the {\it Heschel} bands from the model of  \citet{OssenkopfHenning1994}
with a $\beta$ value of 1.9.}
\label{kappa}      
\centering               
\begin{tabular}{l c c c c c}  
\hline\hline                
\noalign{\smallskip}
$\lambda$      &  $70\,\mu$m & $160\,\mu$m &  $250\,\mu$m &      $350\,\mu$m & $500\,\mu$m  \\
\noalign{\smallskip}
\hline                
\noalign{\smallskip}
$k_\nu$ [cm$^2$/g]                            & 118.0&     24.8 &     11.6 &      5.9 &  2.9\\
 \noalign{\smallskip}
\hline                        
\noalign{\smallskip}
\end{tabular}
\end{table}

The fit of the SED is obtained by using a black-body leaving as free parameters the temperature $T$ and the 
surface  density $\Sigma\, [{\rm g/cm^2}]$. 
A similar approach to obtain temperature and column density has been applied for the 
analysis of the HOBYS key project by e.g. \citet{Hennemannetal2012} and the Gould Belt key program by e.g. 
\citet{Palmeirimetal2013}. 

The final temperature and density maps are obtained from the {\it Herschel}/PACS and SPIRE color corrected mosaics.
\begin{figure*}
\centering
\includegraphics[width=18cm]{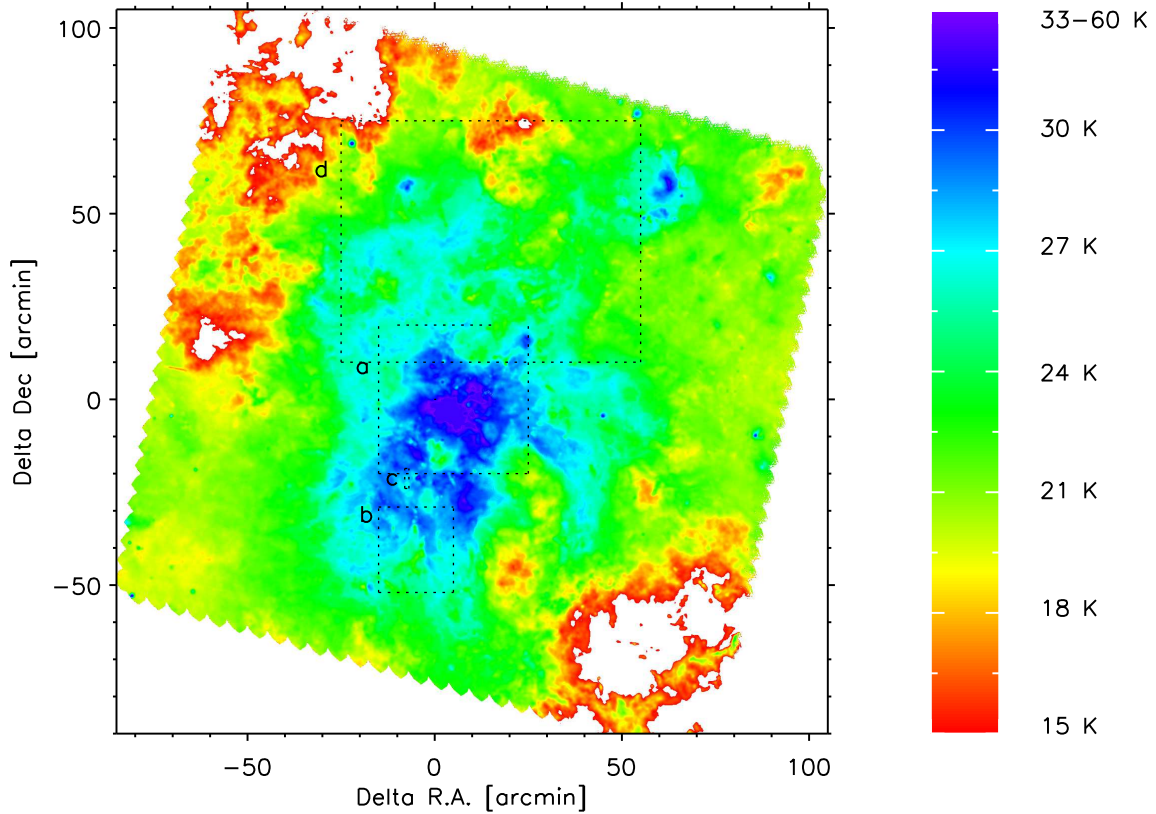}
   \caption{Temperature map of the Carina nebula. 
    The dashed boxes represent: the central region around $\eta$~Car (box `a'), the Southern Pillars region 
    (box `b'), the {\it Treasure Chest Cluster} (box `c'), and the `wave' pattern (box `d'). These regions are 
    analyzed in detail in Figures~\ref{temp_etacar}, \ref{temp_pillars}, \ref{trc}, and \ref{struct_mass}. 
    The center of the figure corresponds to the position $(\alpha_{\rm J2000},\, \delta_{\rm J2000})\, =\, (10^{\rm h}\,45^{\rm m}\,21^{\rm s},\, - 59\degr\,34'\,20'')$.}
     \label{temp}
\end{figure*}

The final temperature map of the CNC is shown in Figure~\ref{temp}.
We find that the average temperature of most of the nebula is about 30~K, ranging between 35-40~K in the 
central clouds and 26~K in the clouds at the edge of the nebula.


In Paper I 
we presented a first color-temperature map 
of the CNC, derived from the ratio of the $70\,\mu$m and $160\,\mu$m images.
The warmer cloud surface temperature is traced with the best resolution, since the convolution 
of the two original mosaics has been done on the PACS 160~$\mu$m.

However, a color temperature may be less well suited for dense clouds 
where much of the 70~$\mu$m - 160~$\mu$m emission comes from the warmer 
cloud surfaces. In this case, the temperature computed by the SED fitting computes 
the beam averaged temperature among the line of sight and it is more sensitive to the 
densest (and thus coolest) central parts of clouds. 

We find that the temperature computed in Paper I 
show values up to $\sim5 \%$ higher than the values with the SED fitting. 
This result was already expected because the color temperature is biased to the warmer cloud 
surface.
 
The second free parameter of the SED fitting is the dust surface density, $\Sigma$.  
 The column density N$_{\rm H}$ is obtained as following
\begin{equation}
     N_{\rm H} = 2  N_{\rm H_2} = \frac{2\cdot \Sigma \cdot R}{m_{\rm H}\cdot\mu_{\rm H_2}}  
\end{equation}
where m$_{\rm H}$ is the hydrogen mass and $\mu_{\rm H_2}$ is the 
mean molecular weight (i.e. 2.8). 
Multiplying by the gas-to-dust mass ratio $R$ (assumed to be 100), we obtain the total column density.

\begin{figure*}
\centering
  \includegraphics[width=18cm]{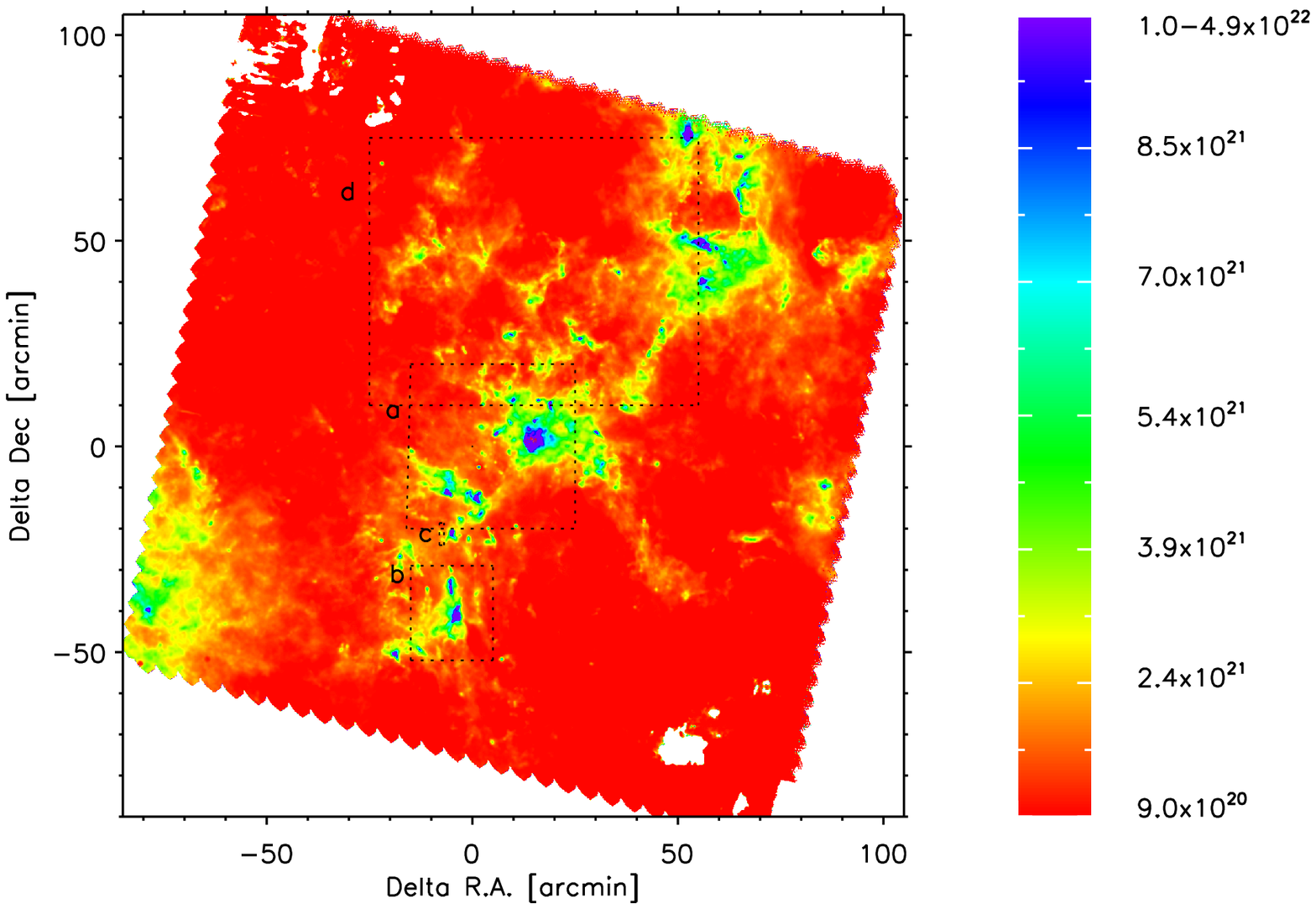}
    \caption{Column density map of the Carina nebula. The color scale is expressed in ${\rm cm^{-2}}$. 
     The dashed boxes represent: the central region around $\eta$~Car (box `a'), the Southern Pillars region 
    (box `b'), the {\it Treasure Chest Cluster} (box `c'), and the `wave' pattern (box `d'). These regions are 
    analyzed in detail in Figures~\ref{temp_etacar}, \ref{temp_pillars}, \ref{trc}, and \ref{struct_mass}. 
    The center of the figure corresponds to the position $(\alpha_{\rm J2000},\, \delta_{\rm J2000})\, =\, (10^{\rm h}\,45^{\rm m}\,21^{\rm s},\, - 59\degr\,34'\,20'')$.}
     \label{sigma}
\end{figure*}

In Figure~\ref{sigma} we show the resulting total 
column density map. 
The median value over the entire CNC (2.3$^\circ\times$2.3$^\circ$) is $1.8 \times 10^{21}\,{\rm cm}^{-2}$, 
while the mean value is $2.3 \times 10^{21}\,{\rm cm}^{-2}$. 

We find column densities up to $5\times10^{22}\,{\rm cm^{-2}}$. 
 From the Southern Pillars, in the South-East direction from Carina, 
there is an almost continuos structure with density of about $3.0-5.5\times10^{21}\,{\rm cm^{-2}}$. 
In the opposite direction, north-east to south-west from the Carina 
Nebula, the density is lower, down to $9\times10^{20}\,{\rm cm^{-2}}$.

In particular the large elongated bubble south-west of
$\eta$~Car is a quite prominent empty region already described in Paper I.

We note that a conservative calibration error of 20\% in the intensity 
causes just 5-10\% uncertainty in the derived temperatures, and
at T$\sim$20~K, a 5\% error in temperature will cause a $\le$10\% error in the derived column 
density.


\subsection{Determination of the UV-flux}

The surfaces of the clouds we see  in the FIR maps are mainly heated by the 
far-UV (FUV) radiation from the hot stars in the complex; the FUV photons
are very efficiently absorbed by the dust grains and thus heat the dust to
temperatures well above the ``general ISM'' cloud temperatures of $\sim 18$~K
\citep[see, e.g.][]{HollenbachTielens1999}. 
Since the heated dust grains cool by emitting
the FIR radiation we can observe with \textit{Herschel}, one can 
estimate the strength of the FUV radiation field from the observed
intensity of the FIR radiation.
This yields important quantitative information about the
local strength of the  radiative feedback from the massive stars.

For a few selected clouds in the CNC, the
strength of the FUV irradiation has been already determined in previous studies.
\citet{Brooksetal2003} investigated the photo-dissociation region (PDR)
at the eastern front of the dense cloud west of Tr~14 and found that
the FUV field is 600 to 10\,000 times stronger than
the so-called Habing field, i.e.~the average intensity of the local Galactic 
FUV ($912\,\AA - 2400\,\AA$) flux of  $1.6\times10^{-3}\,{\rm erg\,cm^{-2}\,s^{-1}}$  \citep[][]{Habing1968}.
\citet{Krameretal2008} derived an independent estimate of the FUV irradiation 
at the same cloud front, finding values of 3400 to 8500 times the Habing field.
For for another cloud south of $\eta$~Car, these authors found irradiation values of
680 to 1360 times the Habing field.

Using the approach described in more detail by \citet{Krameretal2008}, we computed the 
FUV radiation field for all clouds in the CNC from the observed intensity of the
FIR radiation in our \textit{Herschel} maps.
For this, we used our PACS 70~$\mu$m and 160~$\mu$m band maps to determine
the total intensity integrated over the 60~$\mu$m to 200~$\mu$m FIR range.
Since these two bands cover the peak of the FIR emission spectrum of the irradiated clouds, 
they constitute a good tracer of the
radiation from the heated cloud surfaces, and also provide very good angular resolution. 
We did not include the longer wavelength SPIRE bands, firstly because 
the angular resolution in these bands is considerably worse, and secondly
because the emission at $\ga 250\,\mu$m is probably dominated by the thermal emission
from cold dense clumps, that are well shielded from the ambient FUV field;
the emission from these very dense and cold cloud structures does thus not directly trace 
the FUV irradiation.
In any case, the contribution of the SPIRE bands to the integrated intensity would be 
quite small.

To determine the total  $60 - 200$~$\mu$m FIR intensity, $I_{\rm FIR}$, we
first smoothed the PACS 70~$\mu$m map  to the angular resolution of the PACS 160~$\mu$m map
and created maps with a pixel size of $4.5''$ (corresponding to 0.049~pc). 
For each pixel in this map, we
then multiplied the specific intensity in the two bands by the bandwidth
($60 - 80\,\mu$m and $125 - 200\,\mu$m; see PACS MANUAL\footnote{http://herschel.esac.esa.int/Docs/PACS/html/pacs\_om.html}).
In order to fill the gap from $80\,\mu$m to  $125\,\mu$m, we took the average of the
70~$\mu$m and 160~$\mu$m intensities and multiplied it by the corresponding band width.
The resulting sum is a measure of the total  $60 - 200$~$\mu$m FIR intensity ($I_{\rm FIR}$).

Following the arguments of \citet{Krameretal2008}, the FUV field strength
can then be computed via the relation
 \begin{equation}
G_0 = \frac{4\pi\,I_{\rm FIR}}{1.6\times10^{-3}\,{\rm erg\,cm^{-2}\,s^{-1}}}  
\end{equation}
where the denominator contains the Habing field.


\begin{figure*}
\centering
\includegraphics[width=16cm]{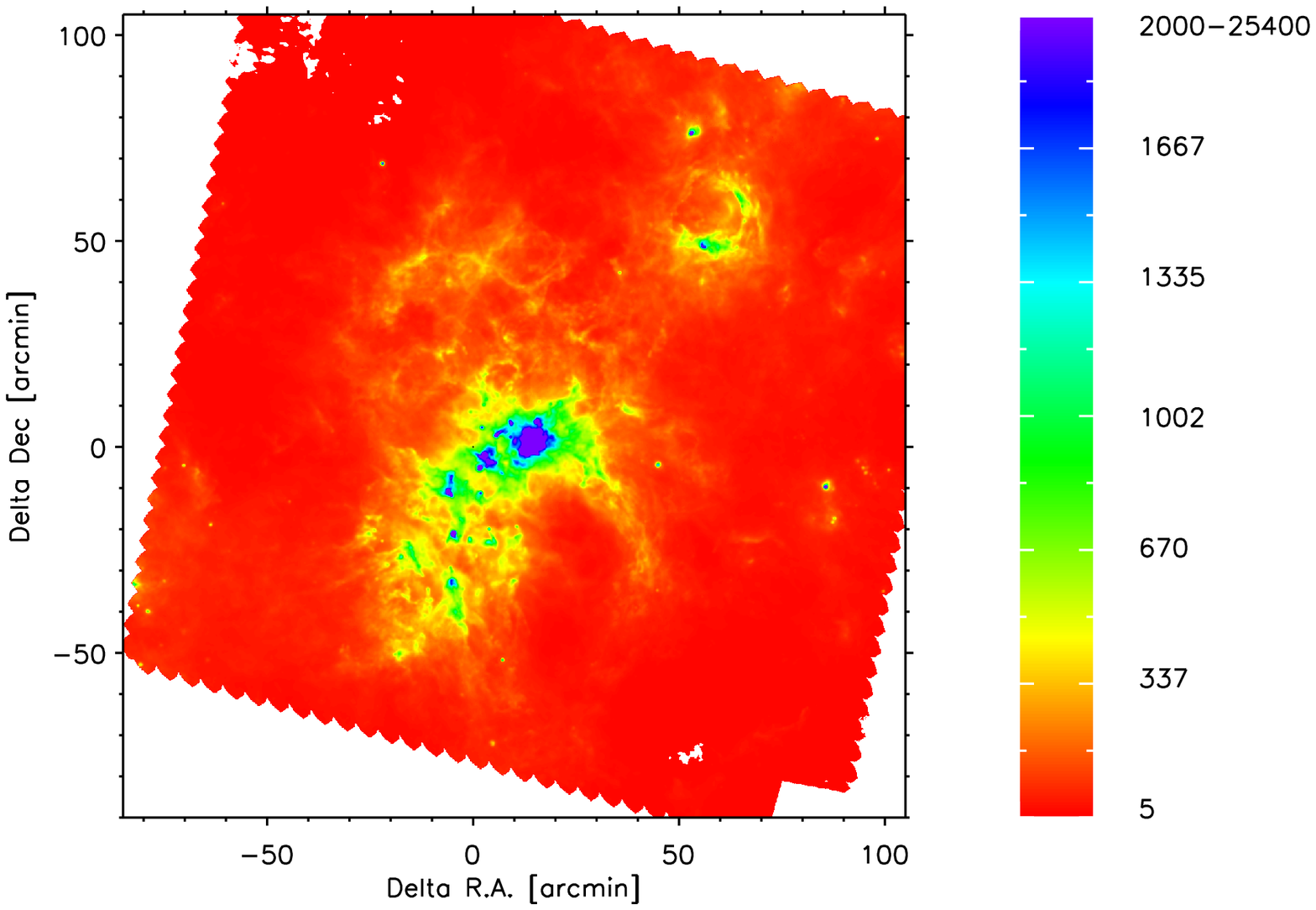}
  \caption{Far-ultraviolet (FUV) flux $G_0$ (in units of the Habing field) of the Carina Nebula Complex . 
   The center of the figure corresponds to the position $(\alpha_{\rm J2000},\, \delta_{\rm J2000})\, =\, (10^{\rm h}\,45^{\rm m}\,21^{\rm s},\, - 59\degr\,34'\,20'')$.}
     \label{uvflux}
\end{figure*}

The resulting map of the derived FUV field in units of the Habing field 
is shown in Figure~\ref{uvflux}. In the enlargements of the FUV map, shown in Figures~\ref{temp_etacar}, 
\ref{temp_pillars}, and \ref{trc}, we also overplotted the positions of the known high-mass stars in the CNC. 
For the individual clouds already studied by \citet{Krameretal2008} and \citet{Brooksetal2003},
our values agree well with the previous, independent determinations by these
authors.

Our map shows that nearly all cloud surfaces in the CNC
are irradiated with $G_0$ values above $\sim 300$, 
 which seems reasonable for a 
region containing  such a large number of OB stars \citep[][]{Hollenbachetal1991}. 
Particularly high G$_0$ values, between $\approx 3000$ and $\approx 10\,000$, 
are found for the massive cloud to the west of Tr~14, 
Values above 3000 are found in the Keyhole Nebula,
several clouds to the north and west of Tr~14,
and a pillar south-east of $\eta$~Car.

It is interesting to compare our map to the results of \citet{Smith2006}, who computed an 
estimate of the radial dependence of the average FUV radiation field 
from the spectral type of each high mass star in the central clusters 
(i.e. Tr~14, Tr~15 and Tr~16). At distances larger than 10~pc from the center of the Nebula,  
the ionizing source can be treated as a point-like source, and the ionizing flux drops as $r^{-2}$.  
The local radiation within few parsec from each high mass cluster is found to be more important 
than the cumulative effect from all the clusters. The FUV radiation computed by \citet{Smith2006} is 
consistent with the values we found: between 100 and 1000 at distances larger than 10 pc from the 
center, and values up to 10$^4$-10$^5$ near the individual clusters. 

In the Southern Pillars area, which is further away from the 
hot stars in the central clusters, 
we find G$_0$ values around 1000 and up to 2000 along the pillar surfaces.

In most parts of our map area, the spatial distribution of the derived FUV intensities correlates
well with the cloud temperatures.
However, in some locations, clear differences can be seen.
One prominent case is the Gum~31 region, in the north-western part of our maps
\citep[see][for a comprehensive study of this region]{Ohlendorfetal2013}.
While the peak of the cloud temperature is found in the diffuse gas around
the OB stars in the central cluster NGC~3324, the
highest FUV intensities are found at the western and southern rim of the
bubble surrounding NGC~3324.

 \section{Detailed structure of the clouds in the Carina Nebula Complex}
 \label{clouds}
 In this section we characterize particularly interesting  small-scale structures in the CNC which are resolved in the 
 {\it Herschel} maps.
In particular we will analyze the central part of the CNC, the Southern Pillars region and the Treasure Chest 
Cluster . 
\subsection{Central part}
\begin{figure*}
\centering
\includegraphics[width=6.5cm]{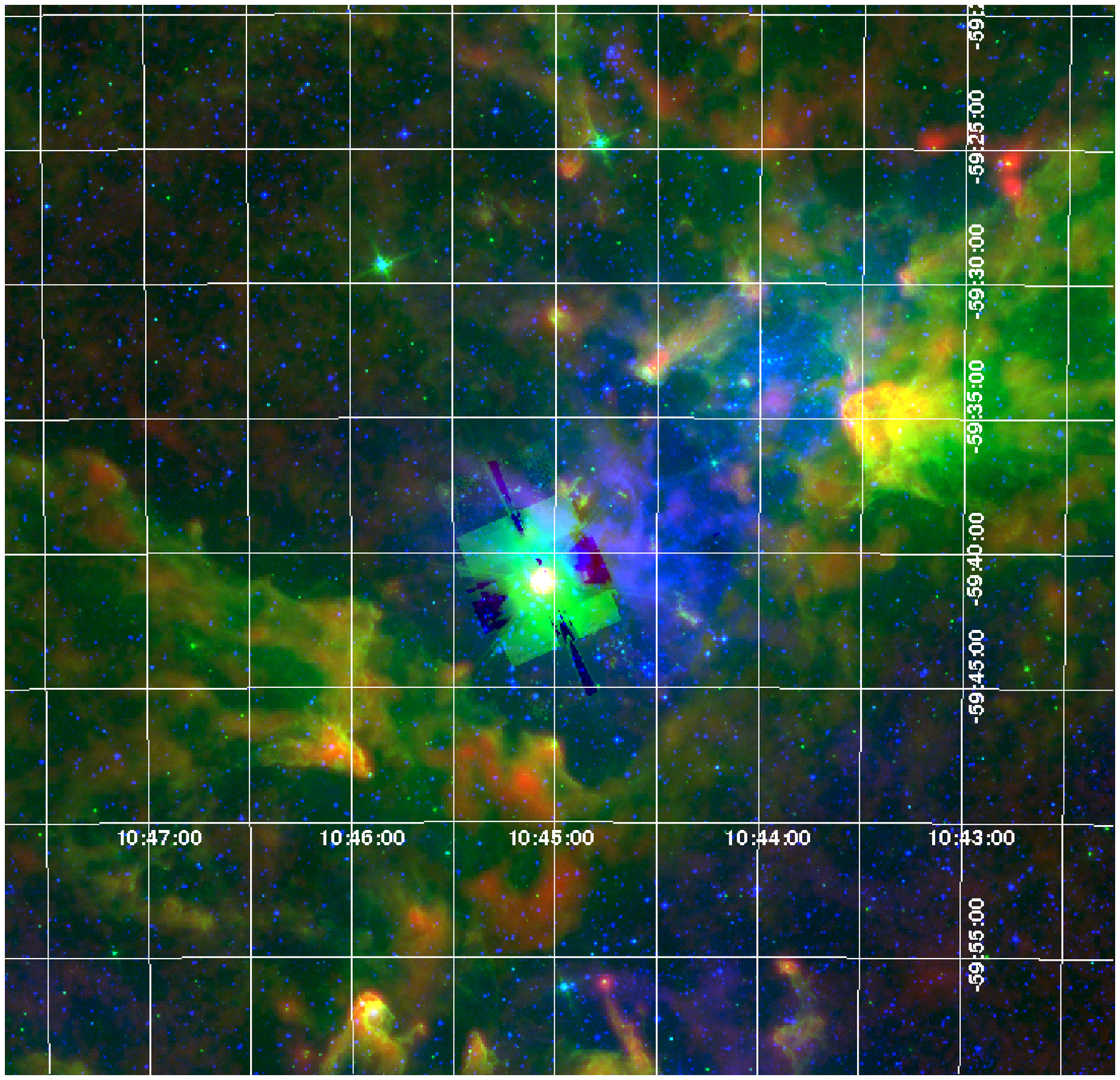}
\includegraphics[bb=0 0 525 396,width=9cm]{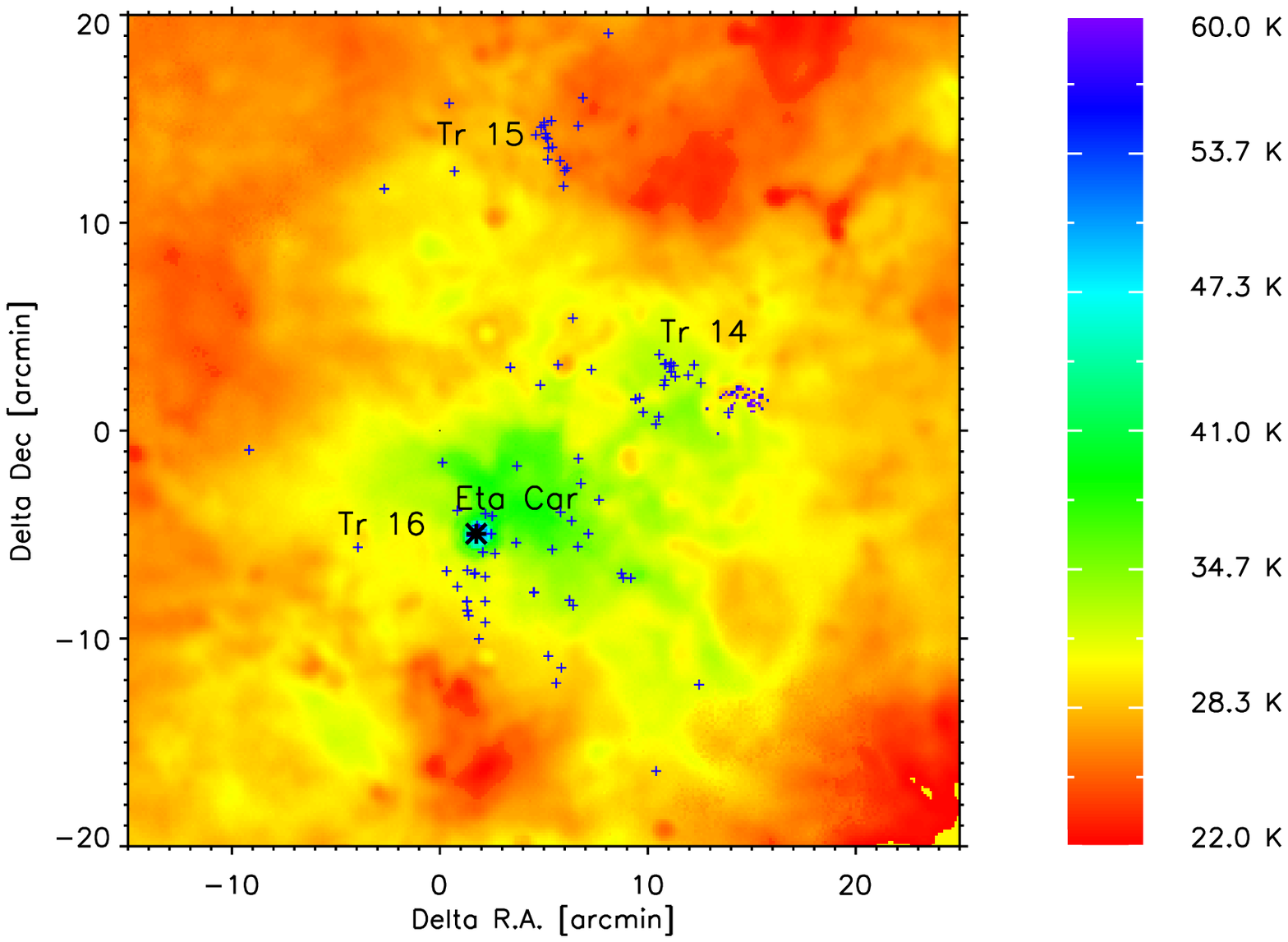}
\includegraphics[bb=0 0 525 396,width=9cm]{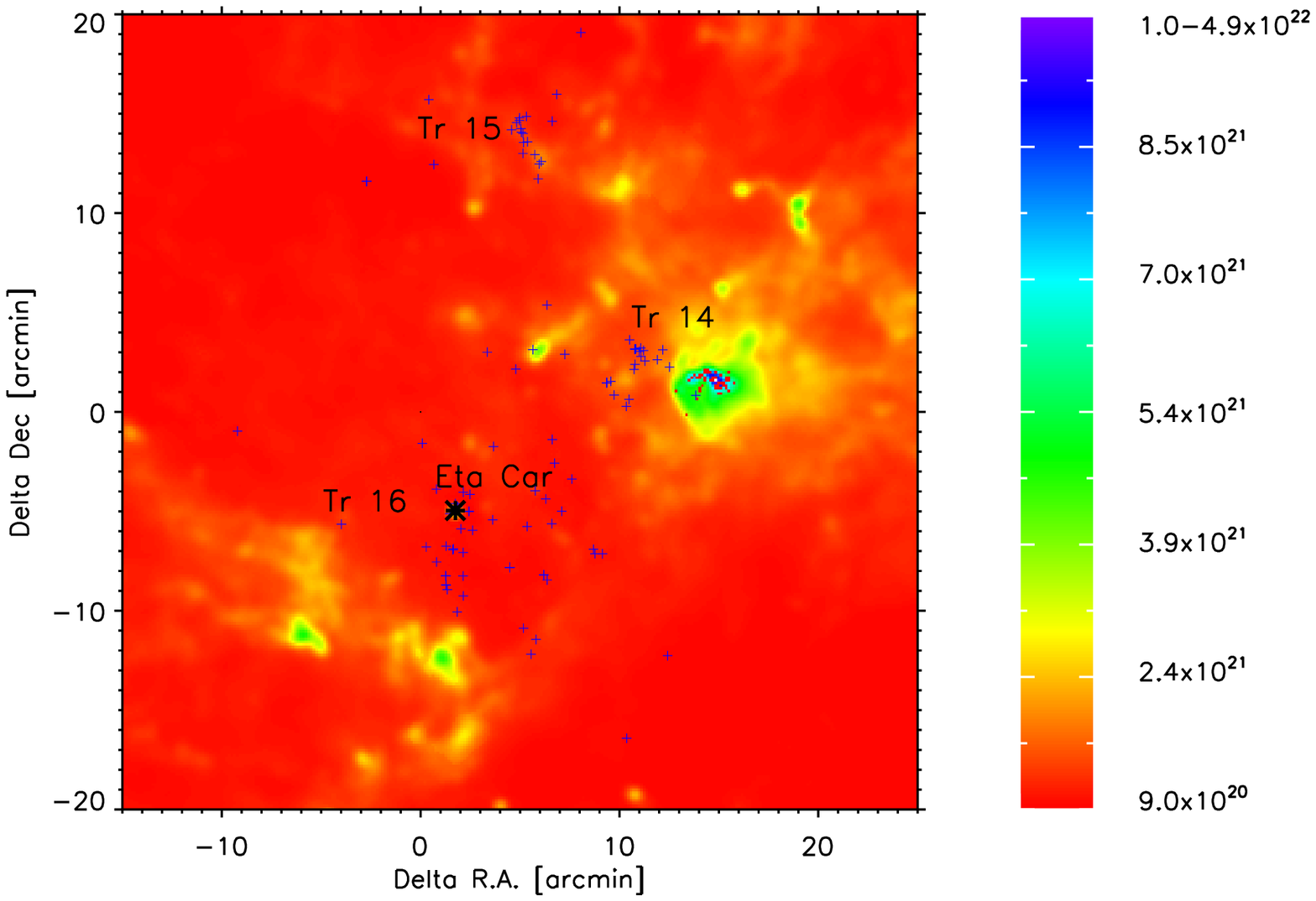}
\includegraphics[bb=0 0 525 396,width=9cm]{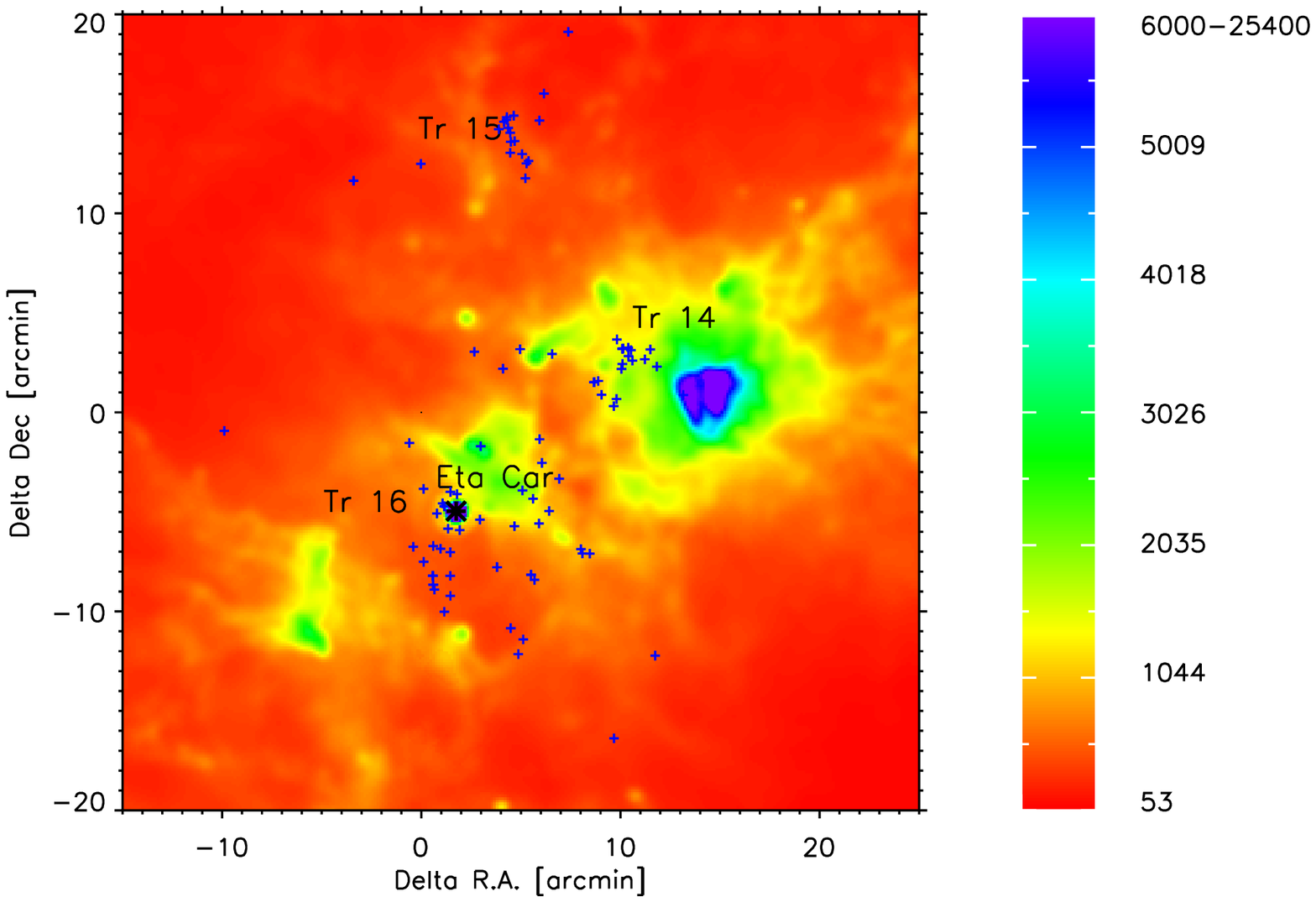}
 \caption{{\it Upper left:} Composite optical DSS (in blue), {\it Spitzer} IRAC4 (in green), and LABOCA (in red) image of the 
  central part of the Carina Nebula Complex around $\eta$~Car (box `a' in Fig.~\ref{temp} and \ref{sigma}). 
 {\it Upper right:} Temperature map of the central part of the Carina Nebula Complex. 
   {\it Lower: } Column density map, expressed in ${\rm cm^{-2}}$ ({\it left}) and Far-ultraviolet (FUV) flux $G_0$ (in units of the 
   Habing field)  ({\it right}) of the  
   enlargement around $\eta$~Car. 
   The blue crosses represent the positions of the high mass stars in this region and the black star the position of 
   $\eta$~Car.}
     \label{temp_etacar}
\end{figure*}

 An enlargement of a $23$\,pc\,$\times\,26$\,pc region around $\eta$~Car is shown 
in Figure~\ref{temp_etacar}. Our \textit{Herschel} maps provide 
sufficient angular resolution to investigate  the 
 temperature and column density structure of the clouds around $\eta$~Car in detail.

In the southern direction from $\eta$~Car a dense molecular cloud ridge is resolved in the 
IRAC image and column density map.  
The temperature is not homogeneous around $\eta$~Car: in the North-West direction from  $\eta$~Car the 
temperature is higher, ranging between 35 and 45~K; in the 
South-East direction the temperature is lower, between 25 and 35~K. 

In this region, there are the young 
clusters, Trumpler 14, 15 and 16, which host about 80\% of the high mass stars of the entire complex.  
This is also the hottest region of the nebula with temperatures ranging between 30 and 50~K (excluding the 
position of $\eta$~Car itself which reaches values of 60~K).

In particular, the clouds in and around the stellar cluster Tr~16, which hosts most of the high mass 
stars in the nebula including $\eta$~Car, have temperatures between 29 and 48~K.
 Tr~14, which hosts 20 O-type stars,  shows cloud
temperatures of 30--33~K,
 and Tr~15, with 6  O-type stars, displays a lower temperature of 26--28~K. 
The local temperature of the cloud is hence related to the number of high mass stars. 

The column density map of the  molecular cloud around the clusters  shows low values of about 
 $9.0\times10^{20}\,{\rm cm^{\rm -2}}$. 
The molecular cloud structures at the edges of the clusters are colder ($\sim23$\,K) and 
denser $\sim3.9\times10^{21}\,{\rm cm^{\rm -2}}$ compared to the central part. 
The most prominent cloud West of Tr~14 has a temperature of about 30~K and a decrease 
in density from the inner to the edge part. 

\subsection{ The Southern Pillars}
\label{SP}
\begin{figure*}
\centering
\includegraphics[width=6.5cm]{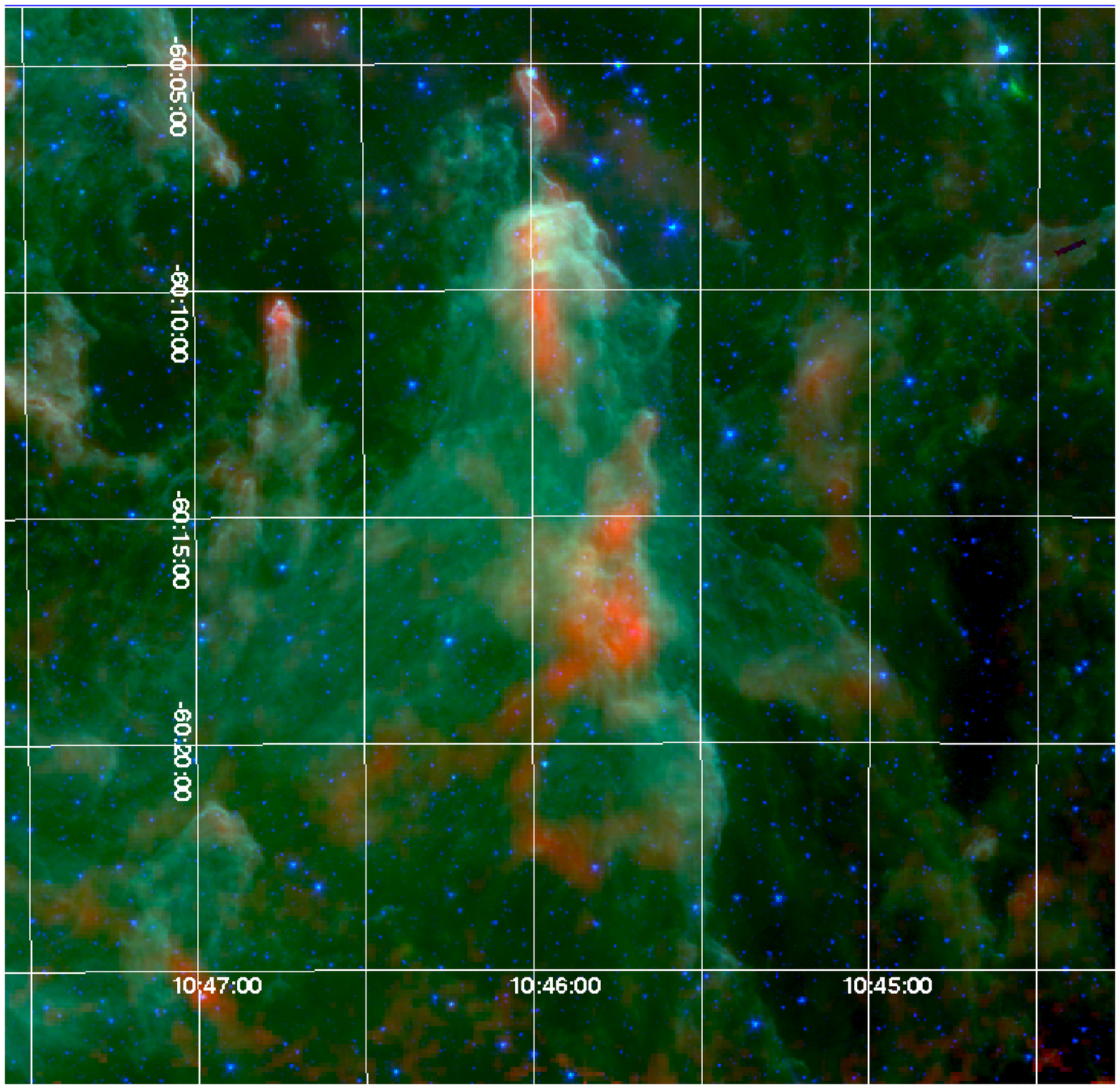}
\includegraphics[bb=0 0 520 396,width=9cm]{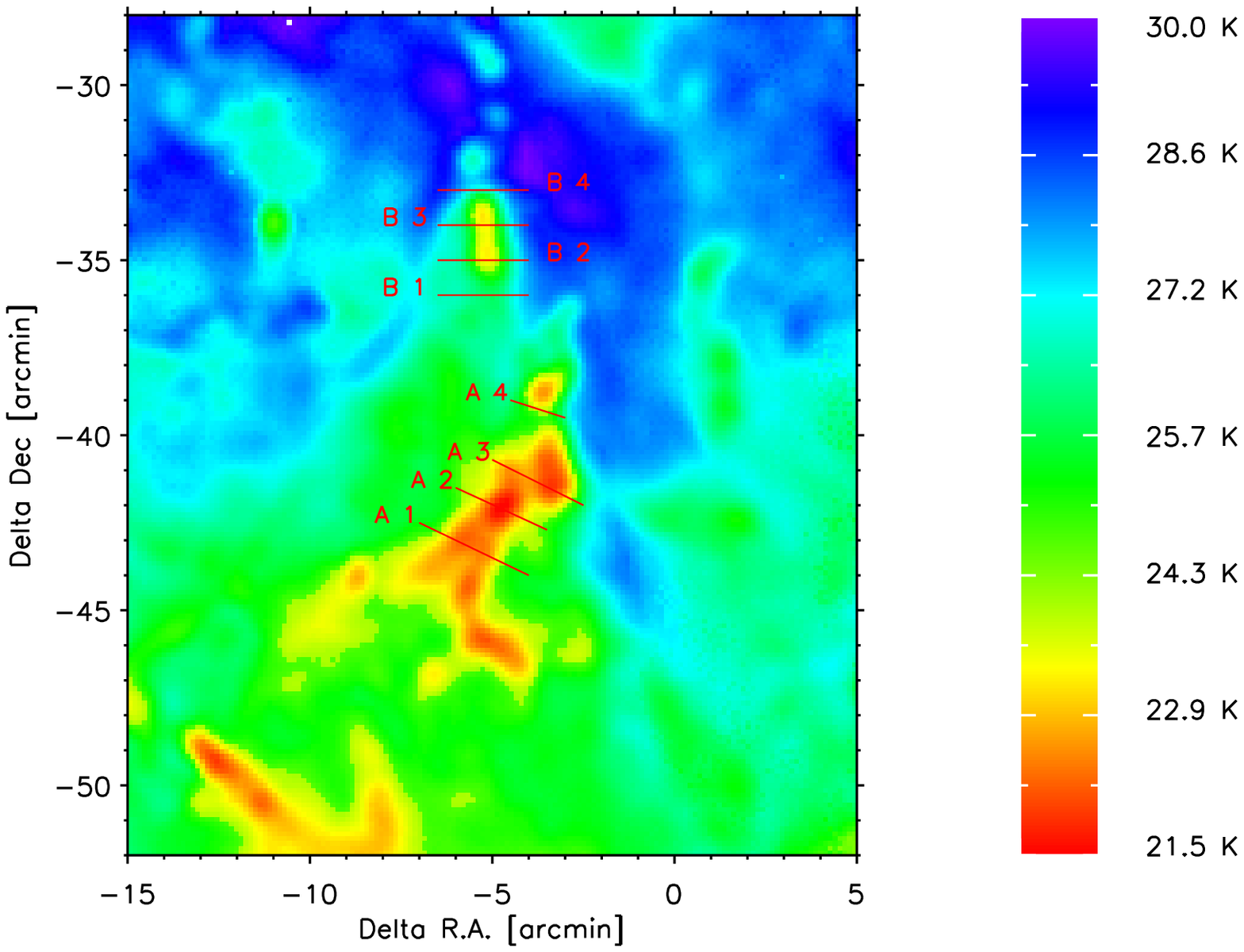}
\includegraphics[bb=0 0 520 396,width=9cm]{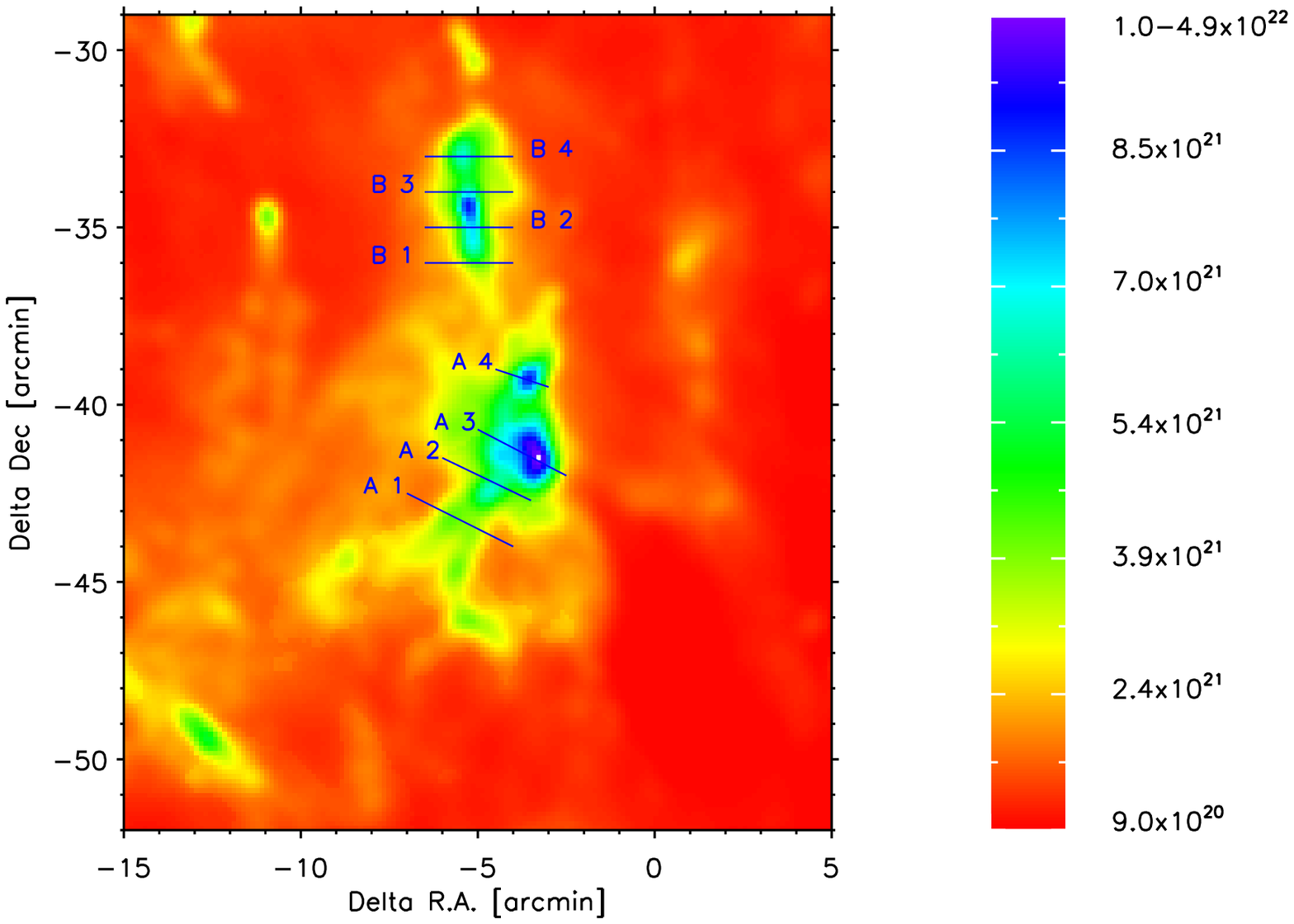}
\includegraphics[bb=0 0 520 396,width=9cm]{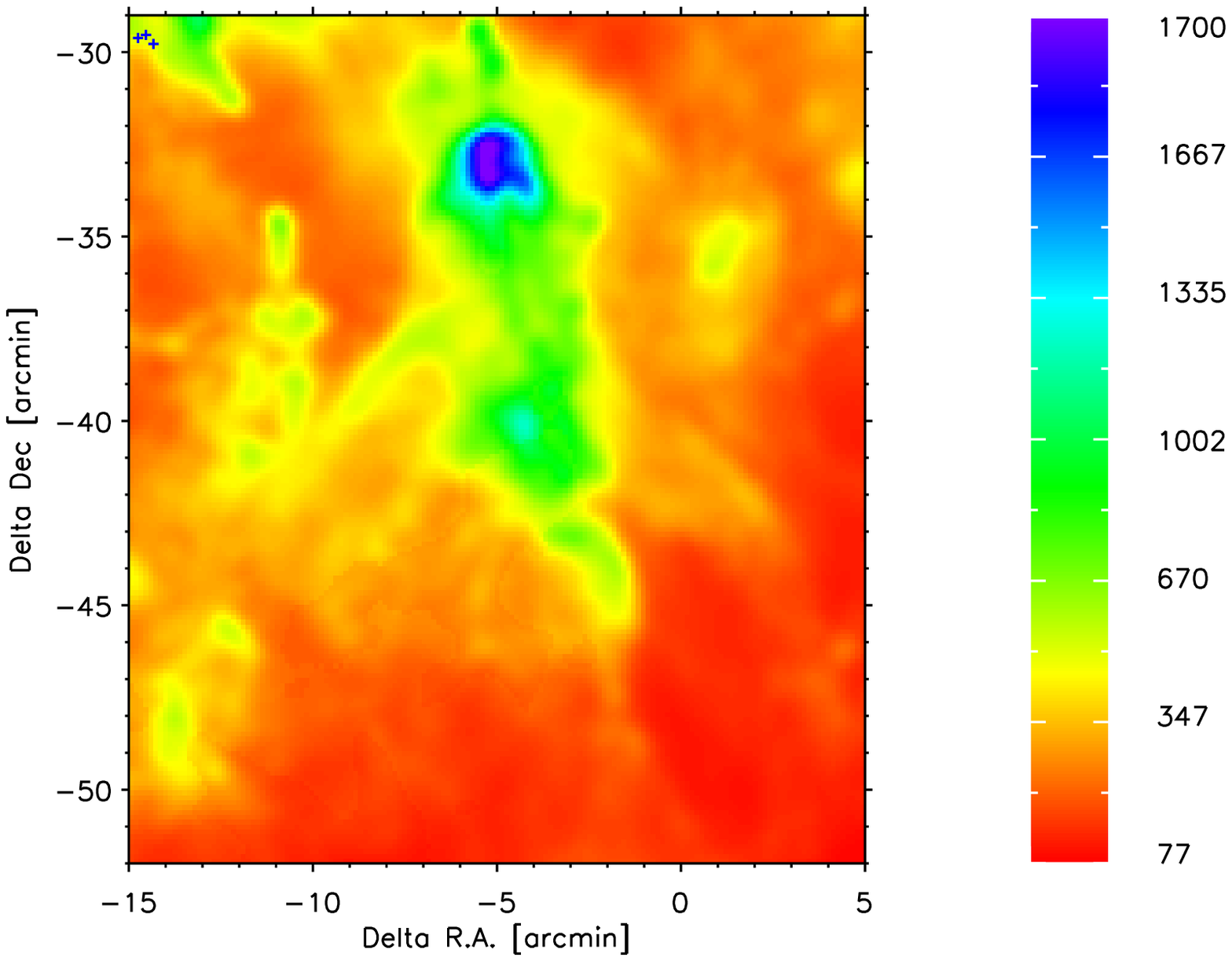}
   \caption{ {\it Upper left:} Composite image  (LABOCA~870~$\mu$m, {\it Spitzer}~3.6 and 8.0~$\mu$m) 
   of the giant pillar in the Southern Pillars region (box `b' in Fig.~\ref{temp} and \ref{sigma}). 
  Temperature map ({\it upper right}), 
   Column density map, expressed in ${\rm cm^{-2}}$ ({\it lower left}) 
   and Far-ultraviolet (FUV) flux $G_0$ (in units of the Habing field) 
     ({\it lower right}) of the enlargement of the region. 
   The solid lines highlight the cut lines of the prominent pillars for which temperature and 
    column density are shown in  
   Figure~\ref{temp_prof_pillars}.}
     \label{temp_pillars}
\end{figure*}

In the Southern Pillars region of the CNC, the temperature ranges between 21~K and 30~K. 
For a few of the most prominent pillars, we analyze the temperature and density profiles in detail.
In Figure\,\ref{temp_pillars} an image of the Southern Pillars is compared with the temperature and 
column density maps. 
All the structures in the composite image in Figure~\ref{temp_pillars} are reproduced in the density map. 
The two most prominent pillars (Pillar~A and Pillar~B in Fig.~\ref{temp_pillars}) have a projected angular 
separation of about 4$''$. In the East direction from 
Pillar~A and Pillar~B another pillar is evident in the composite image of Figure~\ref{temp_pillars}. This is only 
barely visible in the density map, while in the temperature map only a slightly colder temperature compared to the 
local environment is seen at the head of the pillar. 
The projected size of both, Pillar~A and Pillar~B, is about 5~pc. 
The head of the pillars points in the direction of the central stellar cluster Tr~16.   
Their column densities range  from $\sim3.0\times10^{21}\,{\rm cm^{\rm -2}}$
 at the edge of the pillar structure, 
up to $\sim10\times10^{21}\,{\rm cm^{\rm -2}}$ at the peak of density in the central part of Pillar~B. 
In order to analyze the temperature and density profiles along the pillars, we show the run of temperature
and column density along cuts perpendicularly to their main axis, spaced by 
$\sim$1$\arcmin$ from the bottom to their head.  
While the temperature profile always decreases from the edge to the center of the pillar, the density profile 
increases and the maximum of the density in the pillar's center corresponds to the minimum temperature. 
The temperature varies by  2~K according to their position on the pillars: cooler at the bottom of the pillar and 
hotter at their head. 
The Pillar~A is found to be about 2~K cooler than Pillar~B. 
A possible explanation is that Pillar~B is closer to the stellar cluster Bochum~11 which harbours 5 high-mass stars. 
The temperature and density profile of Pillar~A suggests that it consists of two partly resolved pillars: the first 
and third sections show a  double peak in both temperature and density profiles, while the second and fourth 
sections show only one peak which traces one of the two here unresolved pillars. 

\begin{figure*}
\centering
\includegraphics[width=18cm]{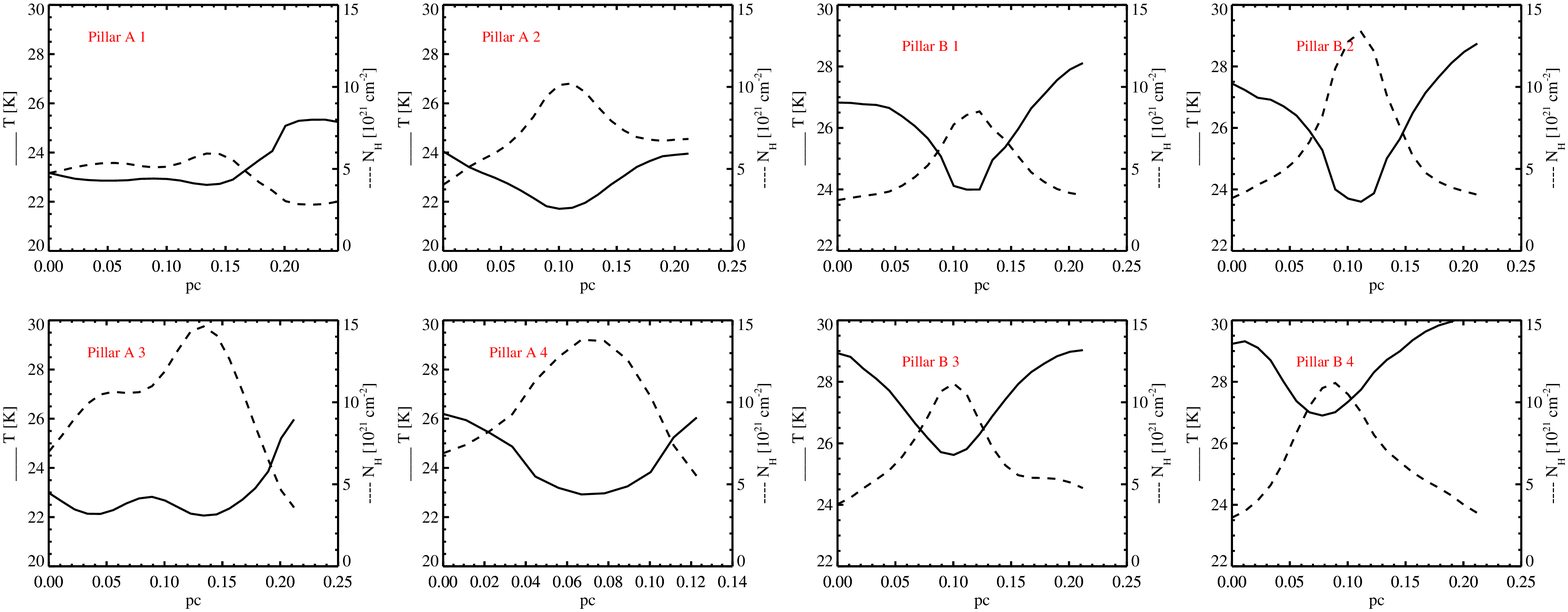}
   \caption{ The plots show the 
   	temperature and column density profiles among the cuts (from left to the right, i.e. in East to West direction) 
	of the prominent pillars highlighted in 
	Figure~\ref{temp_pillars}. The solid lines represent the temperature profiles and dashed lines the column 
	 density. }
     \label{temp_prof_pillars}
\end{figure*}

Integrating our column density map over the spatial extent of the
entire ``giant pillar'' (i.e.~including both, Pillar A and Pillar B),
we derive a total mass of $\approx 9350\,M_\odot$ for the
pillar structure.

The density structure of the pillars has been also predicted by theoretical studies, e.g. by  
\citet{Gritschnederetal2010, Tremblinetal2012}.
In the simulations of \citet{Gritschnederetal2010} {\it Mach 5}\footnote{Details of the {\it Mach 5} simulations are 
in \citet{Gritschnederetal2009}.}, after 500~kyr the ionisation leads the creation of few 
big pillars. The most prominent pillar created by \citet{Gritschnederetal2010} has a size of 0.4 pc\,$\times$\,0.8 pc, 
with column density up to $1.3 \times 10^{22}~{\rm cm^{-2}}$.  
In \citet{Tremblinetal2012} the formation
of a slightly bigger 1 parsec long  pillar takes place in the {\it Mach 1} simulation between 200 kyr and 700 kyr. 
In particular their non-turbulent simulation  was based on the high curvature curvature of the
dense shell which leads to pillars. 

The sizes of the pillars predicted by both, \citet{Gritschnederetal2010} and  \citet{Tremblinetal2012}, are between four and two 
times smaller compared to the extensions of Pillar~A and Pillar~B. In the simulation of \citet{Tremblinetal2012}
it is interesting to notice that the pillar has two dense heads separated by 0.2~pc. This might explain the double 
peak in the temperature and density profile found in Pillar~A. 
The column density predicted by \citet{Gritschnederetal2010} is in good agreement with the profile of Pillar B.

\subsection{The cloud around the {\it Treasure Chest Cluster}}
\label{TCC}
\begin{figure*}
\centering
\includegraphics[width=7cm]{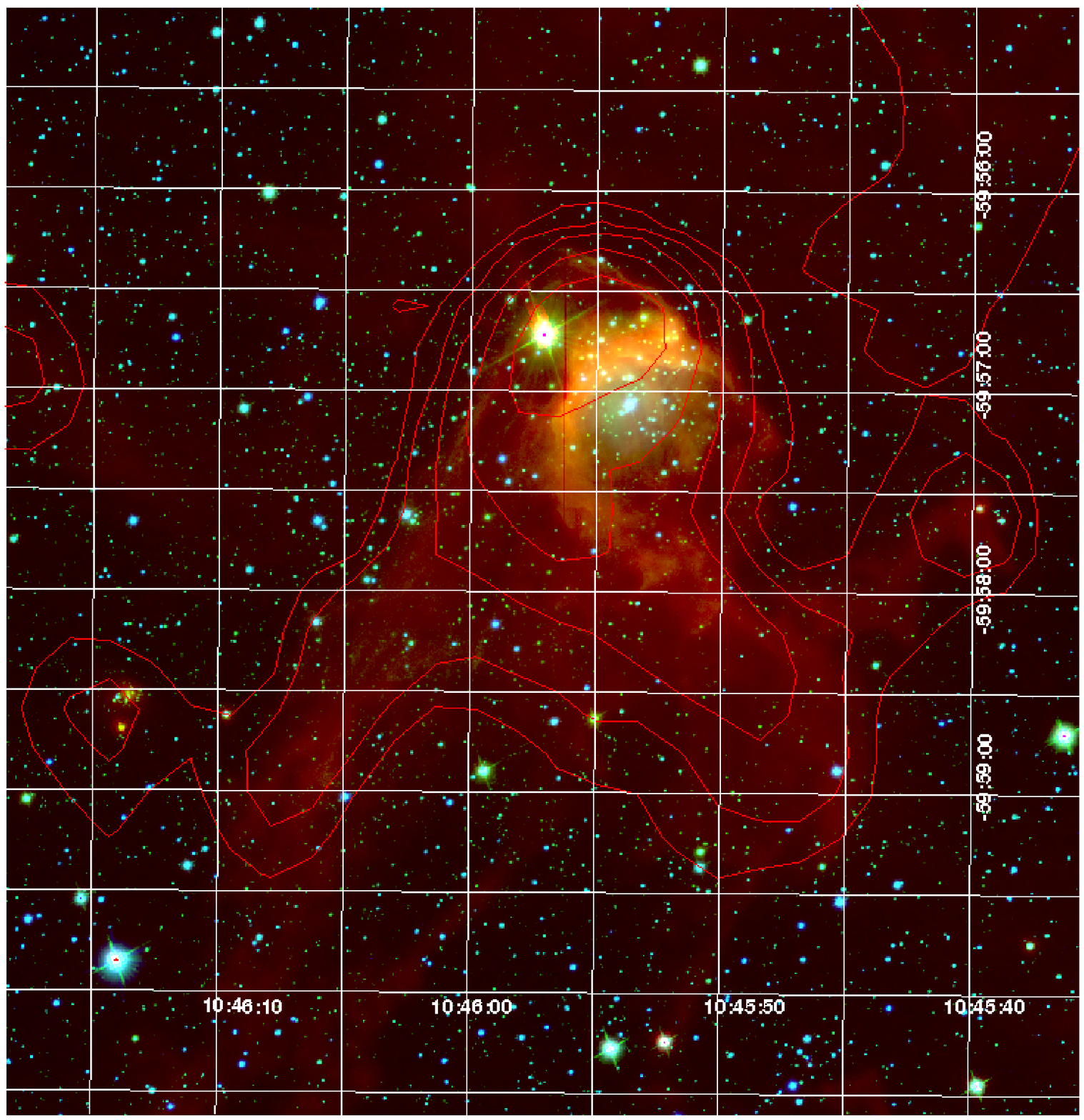}
\includegraphics[bb=0 0 525 396,width=9cm]{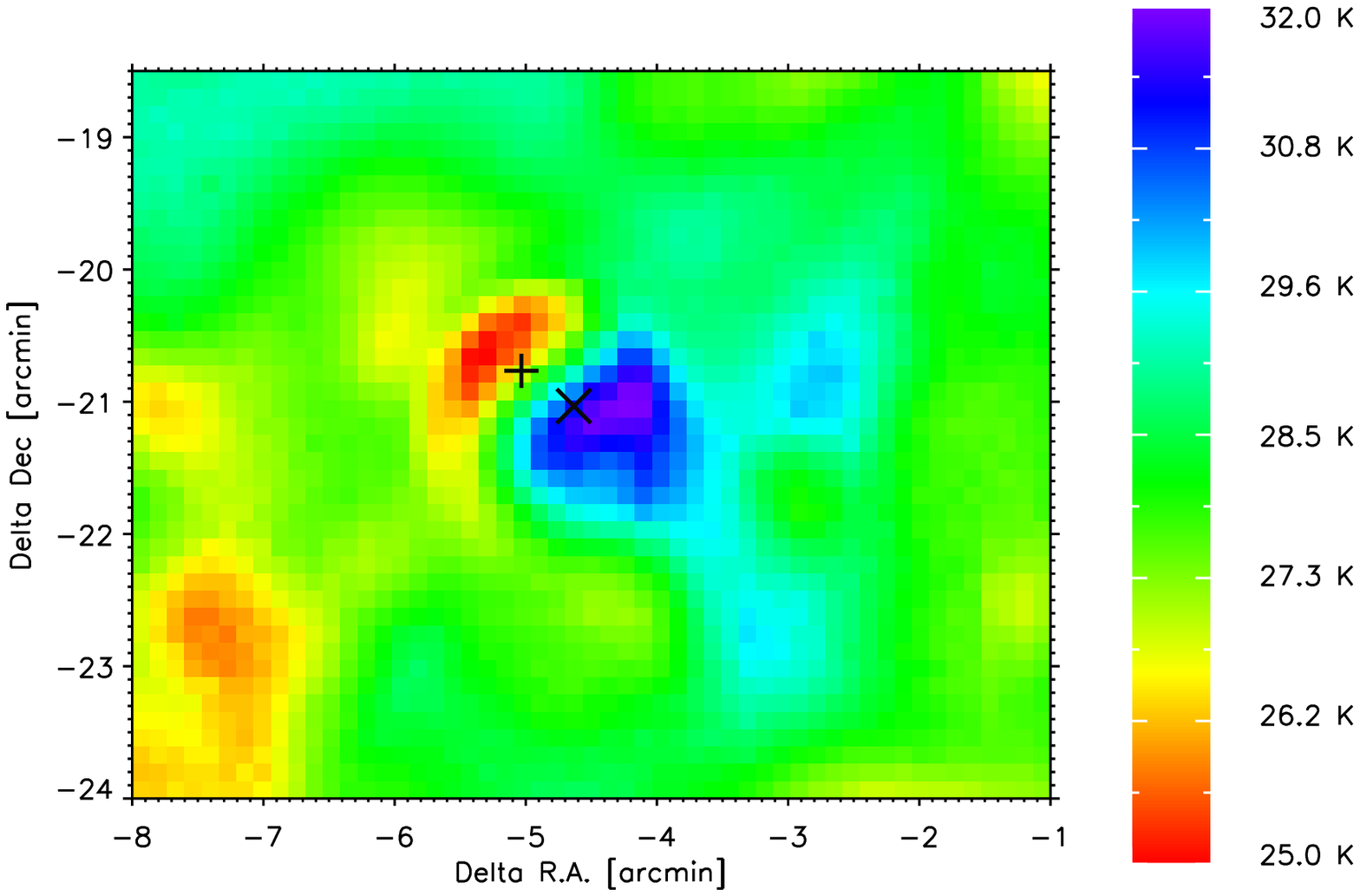}
\includegraphics[bb=0 0 525 396,width=9cm]{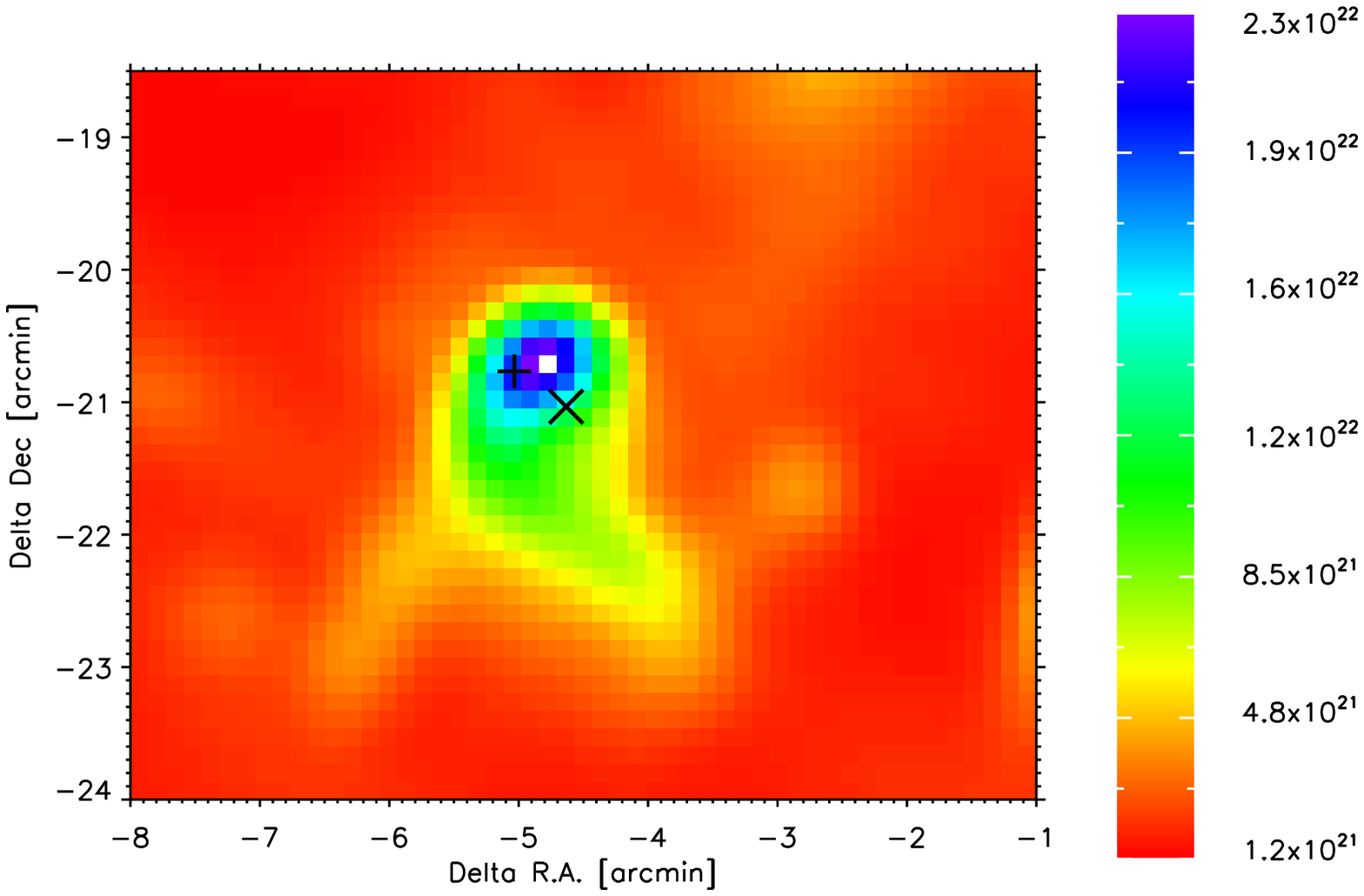}
\includegraphics[bb=0 0 525 396,width=9cm]{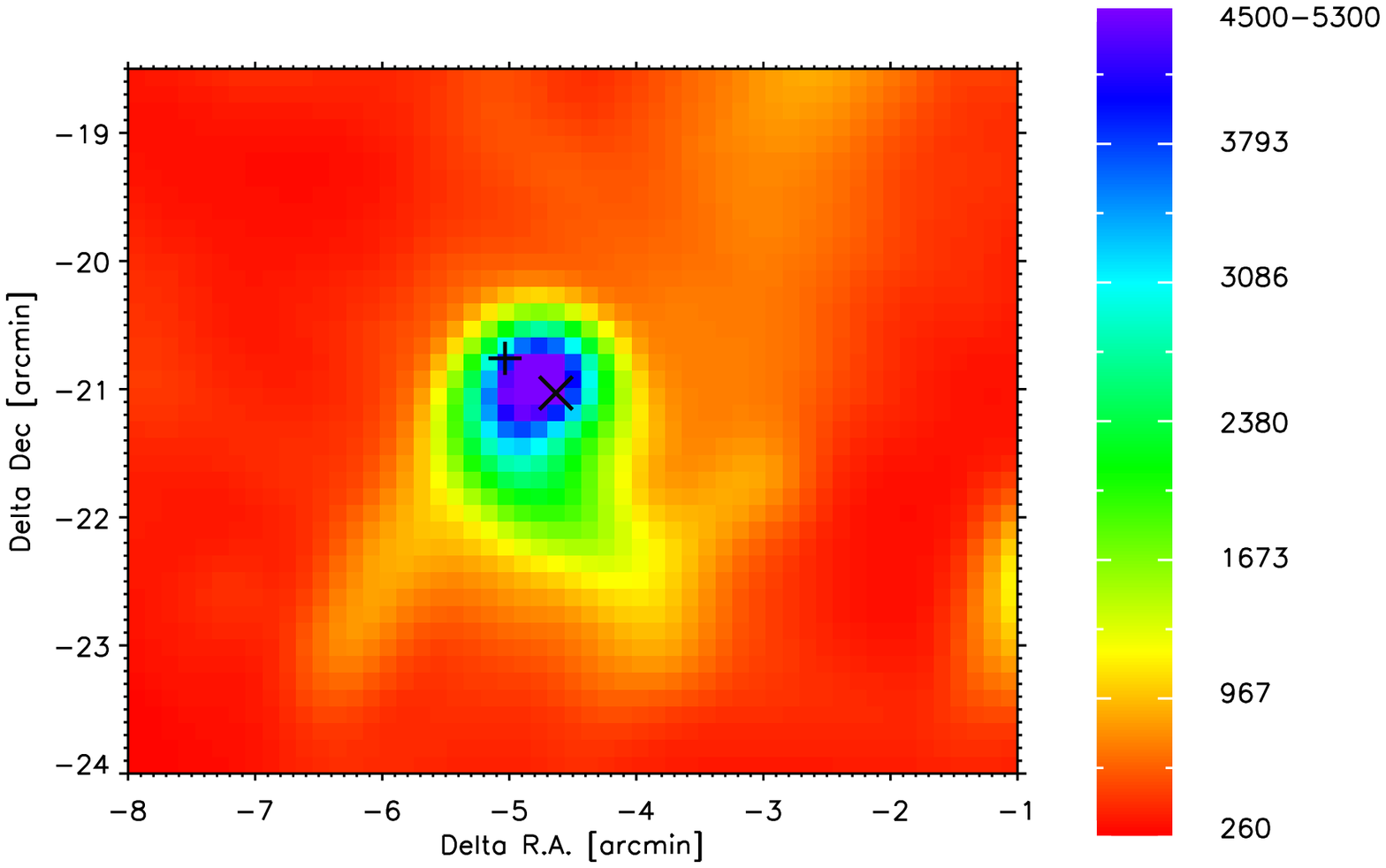}
   \caption{{\it Upper panel:} Composite image of the  {\it Treasure Chest Cluster} (box `c' in Fig.~\ref{temp} 
   and \ref{sigma}): HAWK-I $J$-band image in 
   blue, HAWK-I $K_s$-band image in green, 
   IRAC3 image in red, and the (red) contours of the LABOCA map. 
   In the following panels are shown the temperature ({\it upper right}), column density maps  ({\it lower left}) and 
   Far-ultraviolet (FUV) flux $G_0$ ({\it lower right}). 
   The column density  map is expressed in ${\rm cm^{-2}}$, while the FUV in units of the 
   Habing field.  The black `X' represents the position of the {\it Treasure Chest Cluster}, 
   while the black cross the position of the star  Hen 3-485.	  }
     \label{trc}
\end{figure*}

In the Southern Pillars region, the {\it Treasure Chest Cluster} is the largest and
most prominent embedded star cluster. It is partly embedded in a dense cloud
structure and is thought to be very young, with an estimated 
aged of less than 1~Myr \citep[e.g.][]{Smithetal2005,Preibischetal2011c}. The brightest and
most massive cluster member is the star CPD~$-59\degr\,2661$  (= MJ 632), for 
which a spectral type of B1Ve is listed in the SIMBAD database.
In the near-infrared image (Figure~\ref{trc}), 
the cluster is surrounded by an arc of diffuse nebulosity, which seems to represent
the edge of a small cavity that opens to the south-west.
About $30''$ north-east of the Treasure Chest Cluster, and clearly outside this cavity,
another bright star, Hen~3-485 (= MJ 640, spectral type Bep) is seen.

Our temperature, column density, and FUV intensity maps of the cloud
around the cluster show a very interesting morphology, even if the structures 
are not well resolved. 
The comparison of the near-infrared image with our \textit{Herschel} maps 
shows that the Treasure Chest Cluster is situated near the head
of a pillar-like cloud. Our column density map shows that the cluster is clearly offset
from the density peak ($2.3\times10^{22}\,{\rm cm^{\rm -2}}$) of the pillar. 
The map also reveals a ``kidney''-shaped column density depression near the location of the cluster.
Our temperature map shows colder gas ($T \approx 25$~K) near the column density peak,
and considerably warmer gas ($T \approx 32$~K) at the position and to the south-west of the cluster,
as expected due to the local heating of the dust and gas in the cavity by the cluster stars.
These results support our interpretation that the cluster is located in a small
cavity at the western edge of the pillar, which is open in the south-western direction.

The star Hen~3-485 is seen (in projection) almost exactly at the column density peak of the cloud.
However, its optical brightness and the moderate redenning (the observed color of $B-V = 0.7$
suggest a visual extinction of $A_V \la 3$~mag) show that it is most likely located 
\textit{in front} of the pillar.

Our map of the FUV intensity shows a clear peak just between the two B-stars,
coinciding well with the arc of reflection nebulosity in the near-infrared image.

From our column density map we derive a total mass of $\approx 780\,M_\odot$ for the
pillar-like cloud associated to the Treasure Chest Cluster.

\section{Detection of a ``wave''-like pattern }
\label{wave}

\begin{figure*}
\centering
\includegraphics[bb=0 0 525 396,width=9cm]{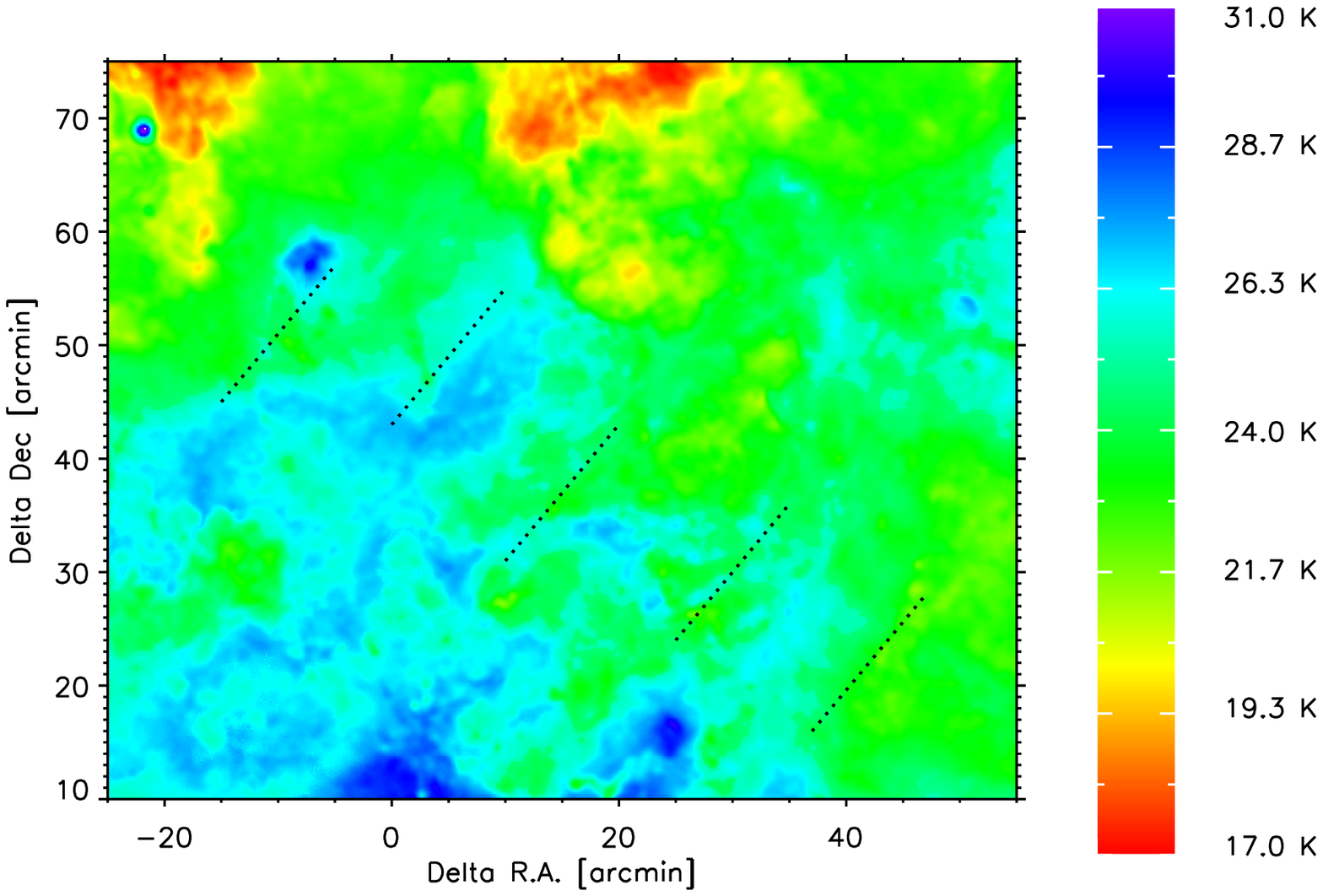}
\includegraphics[bb=0 0 525 396,width=9cm]{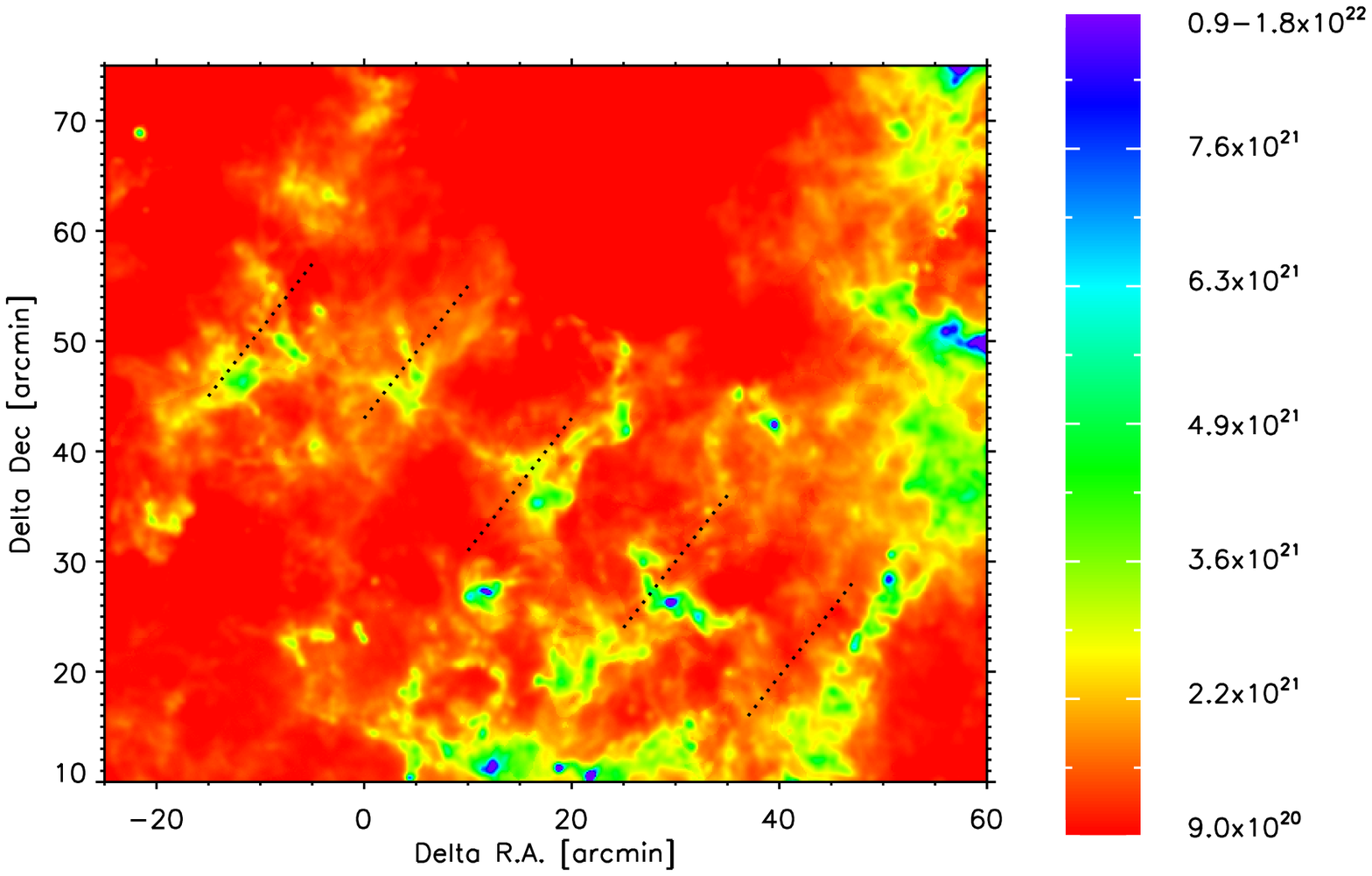}
   \caption{Temperature ({\it left}) and column density, expressed in ${\rm cm^{-2}}$, ({\it right}) map of 
   the north-eastern part of the Carina Nebula Complex (box `d' in Fig.~\ref{temp} and \ref{sigma}). 
    The dashed lines represent a ``wave pattern'' which seems to connect 
   the two main regions of the complex.}
     \label{struct_mass}
\end{figure*}

In the region between the central parts of the Carina Nebula and the Gum~31 region,
in the northern part of our \textit{Herschel} maps, our column density map reveals 
an interesting pattern of quite regularly spaced ``waves'' (see also Figure~2 and 3 of Paper~I). 
It consists of a pattern of parallel lines (see Figure~\ref{struct_mass})
running towards the north-eastern direction.
The wavelength of this pattern is $\approx 14$.$7'$, corresponding to 9.8~pc.
The column density varies from $\sim1.8\times10^{22}\,{\rm cm^{\rm -2}}$ at the maxima to 
$<\,9.0\,\times10^{20}\,{\rm cm^{\rm -2}}$ at the minima.

We first checked whether such a ``wave''-like pattern can be an effect of the observation
mode of Herschel.  
The observations are performed with two scanning directions which are later combined into 
the final map. 
The first scan direction is at 8$^\circ$ and the second one 
is at 100$^\circ$, both from West to North direction. 
The  ``wave''-like pattern is found instead at 40$^\circ$ from the West to North direction, 
which does not correspond to the scan direction. 
We also created a final image from each scan direction  
and the ``wave''-like pattern is present in both of them. For this reason we exclude that this 
is an observational effect.

One possible explanation for this pattern is that we see waves at the surface of the
molecular cloud, similar to the pattern discovered in the Orion Nebula
by \citet{Berneetal2010}.

In the case of the Orion Nebula, the waves have been explained by a Kelvin-Helmholtz 
instability that arises during the expansion of the nebula as
the gas heated and ionized by massive stars is blown over 
pre-existing molecular clouds that
are located a few parsecs away from the center of the cluster, where the
young massive stars are concentrated.

The situation in the  Carina Nebula is similar:
The feedback from the numerous massive stars in the central regions of the
Carina Nebula drives a bipolar bubble \citep[][]{Smithetal2000}.

The location of the observed wave pattern agrees roughly with the northern part
of this bubble system.
The waves are seen at a projected distance of
$\sim 6 - 15$~pc from the massive stars in Tr~16 and Tr~14.
For a comparison to the Orion Nebula case, this larger distance may
be easily compensated by the much larger number of massive stars
in the Carina Nebula and the correspondingly higher level of stellar
feedback. Therefore, the density and velocity of the hot plasma
that is thought to cause the wave pattern in the Orion Nebula
may well be of  a similar order of magnitude  as in the pattern
we see in the Carina Nebula.

Some information about the outflowing material can be obtained from
the \textit{Chandra} X-ray survey in the
context of the \textit{Chandra} Carina Complex Project (CCCP)
\citep[see][]{Townsleyetal2011}.
A detailed analysis of the diffuse X-ray emission, which traces the hot plasma,
is given in \citet{Townsleyetal2011a}.

The largest part of the wave pattern revealed by \textit{Herschel} lies just outside 
the area covered by the  \textit{Chandra} X-ray map.
At  the north-eastern edge of the \textit{Chandra} X-ray map, which
touches the south-western edge of the  wave pattern revealed by \textit{Herschel},
the dominant plasma  components have temperatures ranging from 0.14--0.17~keV
to 0.25--0.27~keV.

The ROSAT image, that covers a larger field-of-view than the \textit{Chandra} map,
shows that the amplitude of the diffuse X-ray emission  in the area of the
wave pattern is rather weak. This suggests that the hot plasma in this region
is not confined, but  streaming outwards.

The most remarkable difference to the waves 
in the Orion Nebula is the substantially longer wavelength of the
wave pattern in the Carina Nebula, $\sim 10$~pc rather than $\sim 0.1$~pc.
According to the models of \citet{Berneetal2010}, this longer wavelength
suggests that a plasma of density $\sim 1\,{\rm cm}^{-3}$ with a 
flow velocity of $\sim 100$~km/s could be causing the instability.
Such values appear quite possible in the case of the Carina Nebula
superbubble.

 \section{Discussion}
\label{discussion}

\subsection{Comparison of dense and diffuse gas}

\begin{figure*}
\centering
\includegraphics[width=14cm]{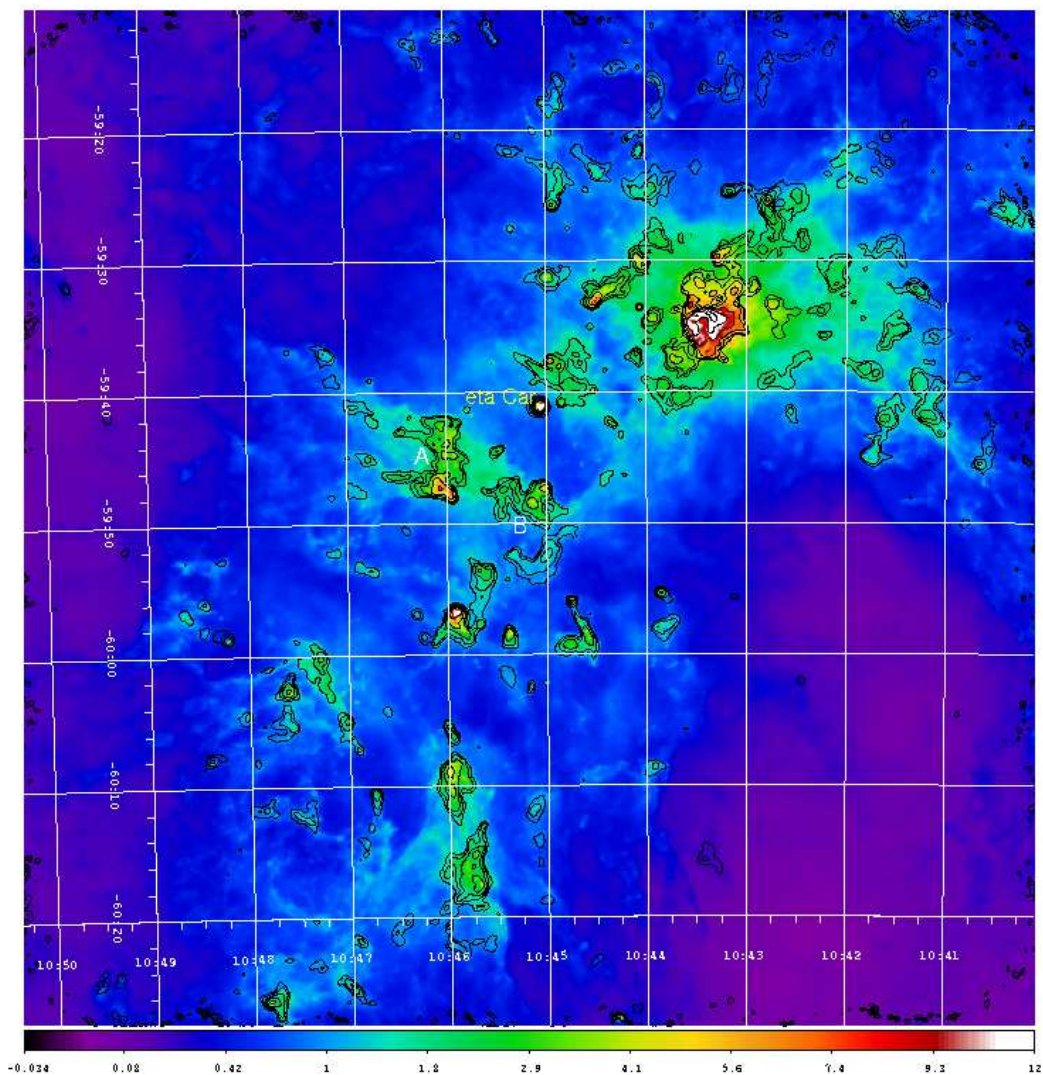}
  \caption{\textit{Herschel}\, $160\,\mu$m map (in Jy/px) with the LABOCA $870\,\mu$m contours 
  overlaid at 0.08,  0.16, 0.32, 0.64, 1.28, 2.56  Jy/beam.  Highlighted is the position of $\eta$~Car.
}
      \label{herschel_laboca}
\end{figure*}

In Fig.~\ref{herschel_laboca} we compare the \textit{Herschel}\, $160\,\mu$m map
to the LABOCA $870\,\mu$m map presented in \citet{Preibischetal2011c}.
This combination has the advantage
that the angular resolution of the  \textit{Herschel}\, $160\,\mu$m map ($\approx 12'' - 16''$) matches the 
angular resolution of the LABOCA map ($18''$) quite well.
The differences in these maps are thus not due to
the difference in angular resolution (as would be the case for the
other \textit{Herschel} maps) but rather trace the differences in the
spatial distributions of the very cold gas and the
somewhat warmer gas.

In most clouds, the cold gas (traced by the sub-mm emission) is considerably more
spatially concentrated than the warmer gas (traced in the \textit{Herschel}\, $160\,\mu$m map).
The most prominent example is the dense cloud west of Tr~14, 
in which the cold dense cloud center is surrounded by a much more extended halo of 
widespread warmer gas.

However, in many of the pillars we find a very different situation.
Most pillars that are seen $\sim 15' - 20'$ to the south-east 
of $\eta$~Car, are not surrounded by significant amounts of extended warm gas.
The spatial extent of the warmer gas matches that of the cold  gas quite well.

The pillars at the southern periphery (more than $\sim 20'$ south of 
$\eta$~Car) show filaments of warmer gas at their base
that seems to stream away from the densest parts of these clouds 
in the direction opposite to the
pillar head, just as expected in the
scenario of cloud dispersal by ionizing irradiation.


A particularly interesting difference in the spatial configuration of the
cold gas versus the warmer gas
is displayed by the cloud structure $\sim 5'$~-~$10'$ to the south and south-east of
$\eta$~Car (that constitutes the ``left wing'' of the {\sf V}-shaped dark cloud).
Here, the cold gas is
highly concentrated in two distinct compact clouds (labeled as
'{\sf A}' and '{\sf B}' in Fig.~\ref{herschel_laboca}), which have a gap
(with a width of about $3'$) between them, just at the position that is
closest (in projection) to $\eta$~Car. In a remarkable contrast,
the configuration of the warmer gas resembles just a 
continuous elongated distribution, with no indication of any gap at this position.
This implies that there is a considerable amount of warmer gas in the region
between these two clumps of cold gas.
 
 \subsection{Dust vs.~molecular gas mass estimates for individual clumps}
 \label{dust2gas}
The area covered by our \textit{Herschel} maps includes 14 of the C$^{18}$O clumps 
detected and characterized in the study of
\citet{Yonekuraetal2005}. Compact clouds  are seen in our \textit{Herschel} maps
at the locations of all of these clumps with exception of C$^{18}$O clump number 10.

In order to derive clump mass estimates from their FIR emission,
we integrated over boxes centered at the cloud positions as seen in our \textit{Herschel} maps 
that correspond to the reported positions of the C$^{18}$O clumps of \citet{Yonekuraetal2005}.
Our box sizes were chosen to be as similar as possible to the sizes given by
\citet{Yonekuraetal2005} (details in Table~\ref{y05}).
Eight of the C$^{18}$O clumps are also located in the field of view of our previous
LABOCA observations \citep[][]{Preibischetal2011d}, from which we already estimated 
clump masses based on their sub-mm fluxes.
In Table~\ref{y05} we summarize the total, gas\,+\,dust, clump masses and temperatures computed using
the different methods.

This comparison shows that the masses computed from the
our \textit{Herschel} maps agree generally quite well to the previous estimates based on the
sub-mm fluxes.
 However, the masses based on FIR and  sub-mm data are always about $2-4$ times lower 
than the masses estimated by \citet{Yonekuraetal2005} from their C$^{18}$O data.

There are several possible explanations for this discrepancy. One major issue are the extrapolation
factors used relating the measured dust or CO molecule emission to the total gas + dust mass.
It is well known that the gas-to-dust ratio, for which we assumed the canonical value of
100, is uncertain by (at least) about a factor of 2.
Concerning the masses derived from CO, it is known that the fractional abundance of
carbon-monoxide may not be constant but depend on the cloud density; thus, the factor used to extrapolate
the CO mass to the total mass can easily vary by factors of 2--3 \citep[see, e.g., discussion in
 ][]{Goldsmithetal2008}.

We also find small, but systematic differences in the derived cloud temperatures.
In most clumps, the dust temperature we derive from our \textit{Herschel} maps is a few Kelvin
higher than the CO gas temperatures derived by \ref{y05}. This is probably related to the fact
to both temperatures represent averages over (i) the corresponding beam area and (ii) 
along the entire line-of-sight though the clumps. 
Since real clumps are not spatially isothermal but have
some kind of (unresolved) density and temperature structure, the different biases of the
different observations result in these different temperature values.

We finally note that the \textit{Herschel} maps show many more dense cold dusty clouds,
which remained undetected in the C$^{18}$O maps.

\begin{table*}
\caption{Mass and temperature derived for the C$^{18}$O clumps from \citet{Yonekuraetal2005}.}
\label{y05}      
\centering               
\begin{tabular}{r c cc r r rrr}  
\hline\hline                
\noalign{\smallskip}
Clump   &\multicolumn{2}{c}{Coordinates}&Size& M$_{\rm FIR}$ \tablefootmark{a} &M$_{\rm Sub-mm}$\tablefootmark{b} & M$_{\rm C^{18}O}$\tablefootmark{c}&T\tablefootmark{a}&T\tablefootmark{c}\\\noalign{\smallskip}
 number &$\alpha_{\rm J2000}$&$\delta_{\rm J2000}$ &[$'$]&  [M$_\odot$]  & [M$_\odot$]  & [M$_\odot$] &[K] &[K] \\
\noalign{\smallskip}
\hline                
\noalign{\smallskip}
1	&$10^{\rm h}\,34^{\rm m}\,23^{\rm s}$ & $- 58\degr\,47'\,00''$ & 6$\times$6 & 420	& -- & 1785		&18-21&16\\
2	&$10^{\rm h}\,36^{\rm m}\,43^{\rm s}$&$-  58\degr\,32'\,39''$&7$\times$12& 1232  	& -- &  3910	&20-26&15\\
4	&$10^{\rm h}\,37^{\rm m}\,41^{\rm s}$&$- 58\degr\,26'\,33''$&4$\times$7& 464 	&  -- & 850		&21-26&20\\
5	&$10^{\rm h}\,38^{\rm m}\,33^{\rm s}$&$- 58\degr\,19'\,15''$&7$\times$7& 269 	& -- &  5185		&25-29&28\\
6	&$10^{\rm h}\,38^{\rm m}\,09^{\rm s}$&$- 58\degr\,45'\,58''$&12$\times$7& 1212 	&  -- & 3145	&23-28&26\\
7	&$10^{\rm h}\,37^{\rm m}\,53^{\rm s}$&$- 58\degr\,55'\,37''$&7$\times$8&885	& -- &  1360		&20-21&17\\
8	&$10^{\rm h}\,41^{\rm m}\,14^{\rm s}$&$- 59\degr\,40'\,50''$&8$\times$7& 2391  &829	& 3145	&22-26&24\\
9	&$10^{\rm h}\,42^{\rm m}\,38^{\rm s}$&$- 59\degr\,26'\,49''$&5$\times$5& 476	& 593&1190		&23-28&21\\
11	&$10^{\rm h}\,43^{\rm m}\,56^{\rm s}$&$-59\degr\,24'\,21''$&5$\times$5& 610	& 312&1190		&23-25&22\\
12	&$10^{\rm h}\,45^{\rm m}\,15^{\rm s}$&$-59\degr\,48'\,22''$&6$\times$5& 882 	&1095& 2210		&23-26&20\\
13	&$10^{\rm h}\,45^{\rm m}\,48^{\rm s}$&$-60\degr\,18'\,09''$&7$\times$9& 1748 	& 1212&3570	&22-27&20\\
14	&$10^{\rm h}\,48^{\rm m}\,14^{\rm s}$&$-59\degr\,58'\,48''$&3$\times$3& 161& 85& 382			&23-26&9\\
15	&$10^{\rm h}\,47^{\rm m}\,50^{\rm s}$&$-60\degr\,26'\,15''$&4$\times$5& 353 	&453& 1700		&20-23&19\\
\noalign{\smallskip}
\hline                        
\noalign{\smallskip}
\end{tabular}
\tablebib{
\tablefoottext{a}{This work; }
\tablefoottext{b}{\citet{Preibischetal2011d} assuming a temperature of 20~K; }
\tablefoottext{c}{\citet{Yonekuraetal2005}, scaled to the distance of 2.3~kpc.}

}
\end{table*}


\subsection{Cloud mass budgets based on different tracers}

In Table~\ref{mass_methods} we summarize the integrated cloud masses for different parts of the
CNC derived from different tracers. 
The masses derived in this paper via fitting of the FIR spectral energy distribution,
and in \citet{Preibischetal2011d} 
 from sub-mm fluxes,  are based on dust as a mass tracer.
The study of \citet{Yonekuraetal2005} gives  mass estimates derived from different CO isotopologues,
and is thus based on gas tracers.

Considering the total cloud complex, we find rather good agreement
between the different mass estimates for specific column density thresholds.
Our \textit{Herschel} based cloud mass estimate for an column density threshold
corresponding to 
$A_V > 1$ is well consistent with the mass estimate based on $^{12}$CO.
This is in good agreement with the standard idea that molecules exist 
inside the shielded regions of dark clouds \citep[][]{Goldsmithetal2008}.
The thresholds in $A_V$ of 0.9 and 7.3 have been chosen as in \citet{Ladaetal2012}. 

Our previous cloud mass estimate based on the LABOCA sub-mm flux \citep{Preibischetal2011d}
agrees well with the mass estimate based on $^{13}$CO.
This fits with the assumption that the LABOCA sub-mm map traces
only dense clouds and that $^{13}$CO is a tracer of relatively dense gas.

The total mass of all individual clumps detected in our LABOCA sub-mm map
\citep[as determined by ][]{Pekruhletal2013} is between the cloud mass estimates based
on $^{13}$CO and C$^{18}$O, in good agreement with the idea
that C$^{18}$O traces denser gas than $^{13}$CO.

\subsection{The strength of the radiative feedback}

As mentioned above, the cloud surfaces in the CNC
are strongly irradiated: we find $G_0$ values generally between $\sim 1000$
and up to $\approx 10\,000$.
These numbers provide important quantitative information on the
level of the radiative feedback from the massive stars on the clouds.

To put these numbers in context, we refer to the compilations  
in \citet{Brooksetal2003} (Tab.~3) and \citet{SchneiderBrooks2004} (Tab.~2), where values of the 
FUV field strengths in photo-dissociation regions (PDRs)
in different star forming regions are listed.
This comparison shows that the FUV field strength at the cloud surfaces in the
CNC is very similar to that measured for PDRs in the 30~Dor cluster, which 
is generally considered as the most nearby extragalactic starburst system.
This result supports the idea that the level of massive star feedback
in the CNC is similar to that found in starburst systems.

The comparison also shows that, sometimes, even higher values
of the FUV field strength are found at specific locations in much less massive 
star forming regions. For example, in the PDR in the Orion KL region, $G_0$ values
up to $10^4 - 10^5$ have been determined \citep[see references in ][]{SchneiderBrooks2004}.
Such high local values are, however, only found in clouds located very close 
(less than 1 parsec) to individual high-mass stars, i.e.~restricted to very small 
cloud surface areas. 
No such extremely high irradiation values are seen in the CNC, because the clouds 
are generally at least a few parsecs away from the massive stars. This is
probably a consequence of the fact that many of the massive stars in the CNC 
are several Myr old and have thus already dispersed the clouds 
in their immediate surroundings (by their high level of feedback).

Despite the fact that some lower-mass star forming regions show
peaks of FUV irradiation that can locally exceed the levels we see in the
CNC, the total amount of hard radiation (and thus the total power available
for feedback) is much higher in the CNC.

The high level of the FUV irradiation may also affect the
circumstellar matter around the young stellar objects in the CNC.
The irradiation causes increased external heating and photoevaporation of the
 circumstellar disks and can disperse the disks within a few Myr.
This process has been directly observed in the so-called
``proplyds'' near the most massive stars in the Orion Nebula Cluster
\citep[][]{Odelletal1993,  Ballyetal2000, Riccietal2008}
and has been theoretically investigated by numerical simulations
\citep[e.g., ][]{RichlingYorke2000, Adamsetal2004, Clarke2007}

Using near-infrared excesses as an indicator of circumstellar disks,
our previous studies of X-ray selected young stars in the different populations
in the CNC actually showed that the infrared excess fractions for the
clusters in the CNC are lower than those typical for
stellar clusters  of similar age with lower levels of irradiation \citep[][]{Preibischetal2011d}.
This is consistent with the idea that the high level of massive star
feedback in the CNC causes faster disk dispersal.
Similar indications have been found in other high-mass star forming regions,
where the disk fractions were also found to be lower
close to high mass stars \citep[e.g.][]{Fangetal2012a, Roccatagliataetal2011}.

\subsection{On the relative importance of massive star feedback and  primordial turbulence}

A fundamental question of star formation theory 
is whether the structure of the clouds in a star forming region
is dominated by primordial turbulence or by massive star feedback.

The recent study of \citet{Schneideretal2010} analyzed \textit{Herschel} observations of the 
Rosette molecular clouds and suggested that
``there is no fundamental difference in the density structure
of low- and high-mass star-forming regions''.
They argue that the primordial turbulent
structure built up during the formation of the cloud, rather than the feedback from 
massive stars, is determining the course of the star formation process in this region.
The current star formation process is concentrated in the most massive
filaments that were created by turbulent gas motions. 
The star-formation can be triggered in the direct interaction zone between 
the HII-region and the molecular cloud but probably not deep into the cloud on a size 
scale of tens of parsecs.

It is interesting to compare the  Rosette molecular cloud to the CNC.
The Rosette cloud has a similar, only slightly lower total cloud mass than the CNC,
and also harbors a rather large population of OB stars.
However \citep[as already pointed out by ][]{SchneiderBrooks2004, Schneideretal1998}, 
the Rosette Molecular Cloud is exposed to a much (about 10 times) weaker UV field
than what we find in the Carina Nebula.
This order-of-magnitude difference in the level of the radiative feedback may explain
the differences in the cloud structure of these two complexes. In strong contrast to
the predominantly filamentary cloud structure of the Rosette region,
most clouds in the CNC show pillar-like shapes. Our  \textit{Herschel} maps
reveal
\citep[as previously suggested on the basis of the \textit{Spitzer} data; ][]{Smithetal2010} 
the systematic and ordered shape and orientation of these numerous pillars, that
point towards clusters of high-mass stars. As discussed in \citet{Gaczkowskietal2013},
the current star formation activity in the CNC is largely concentrated to these pillars.
These results clearly suggest that the cloud structure and the star formation activity
in the CNC is dominated by stellar feedback, rather than random turbulence.

A strong feedback similar to the situation in the CNC has been found in Herschel 
study of the RCW36 bipolar nebula in Vela C by \citet{Minieretal2013}.

%
%

\subsection{Total FIR luminosity}

From our map of the FIR intensity we also computed the total FIR
luminosity in the $60 - 200$~$\mu$m wavelength range,  $L_{\rm FIR, tot}$
by integrating over a $2.3\degr \times 2.3\degr$ region covering the
entire CNC. We find a value of $L_{\rm FIR, tot} = (1.7 \pm 0.3)\times 10^7\,L_\odot$.

This value is somewhat larger than the total infrared luminosity
of $1.2 \times 10^7\,L_\odot$ measured by \citet{SmithBrooks2007} 
from IRAS and MSX data, but this difference can be explained by the
slightly smaller area used in the study of \citet{SmithBrooks2007}, that
excluded the Gum~31 region (see their Fig.~1c).

It is interesting to compare this total FIR luminosity to the total luminosity
of the known massive stellar population in the CNC, as derived by
\citet{Smith2006}: 
Our total FIR luminosity is smaller than the total stellar bolometric luminosity
of $2.5 \times 10^7\,L_\odot$, but 70\% larger than the
total stellar FUV  luminosity of $1.0 \times 10^7\,L_\odot$.
The most likely explanation for this apparent discrepancy of the
total FIR and total  FUV  luminosity is that the current census of massive stars
in the CNC is still incomplete. Direct support for this assumption comes from the
study of \citet{Povichetal2011}, who could identify 94 new candidate OB stars
(with $L_{\rm bol} \ge  10^4\,L_\odot$) among X-ray selected infrared sources in the
Carina Nebula. They found that the true number of OB stars in the CNC
is probably about 50\%, and perhaps up to 100\%, larger than the currently identified OB population.
This incompleteness results from the lack of sufficiently deep wide-field spectroscopic surveys,
that left a significant number of OB stars yet unidentified.

Taking this incompleteness into account, the extrapolated total stellar FUV  luminosity
would match the total FIR luminosity we derived quite well.

\subsection{The Carina Nebula as a link between local and extragalactic star formation}

\
\citet[][]{Ladaetal2012} collected observational data for numerous Galactic star forming regions
and external galaxies
and found a well-defined scaling relation between the star formation rates and
molecular cloud masses above certain column density thresholds.
They expressed the star formation scaling law for these clouds as
\begin{equation}\label{sfr}
{\rm SFR} =  4.6 \times 10^{-8}\,f_{\rm DG}\,M_{G}(M_\sun)\,M_\sun\, {\rm yr}^{-1}
\end{equation}
where $M_G$ is molecular mass measured at a particular extinction
threshold and corrected for the presence of helium and $f_{\rm DG}$ is the
fraction of dense gas.

They suggest that this fundamental relation
holds over a span of mass covering nearly nine orders of magnitude.
Although the exact physical meaning of certain density threshold values is not clear
\citep[see discussion in ][]{BurkertHartmann2012}, 
this empirical relation provides an interesting opportunity to compare different
star formation regions.
In the compilation of observational data of \citet[][]{Ladaetal2012}, 
there is a gap of about four orders of magnitude
(in cloud mass) between the local Galactic clouds and the external galaxies.
The CNC is ideally suited to bridge this gap and to connect Galactic and extragalactic
regions in this investigation.

The cloud masses for the CNC determined with different tracers have been
summarized above.
For the extinction thresholds
of $A_K>0.1$~mag and $A_K>0.8$~mag\footnote{$A_K=0.1$~mag $\simeq$ $A_V=0.9$~mag;  
$A_K=0.8$~mag $\simeq$ $A_V=7.3$~mag.}, as used in the study of \citet[][]{Ladaetal2012}, we
find total cloud masses of $\sim 610\,000\,M_\odot$
and $\sim 23\,000\,M_\odot$ from our \textit{Herschel} data of the entire CNC.

Taking into account the systematic calibration uncertainties of the Herschel fluxes as well as 
 uncertainty of the exact boundaries of the CNC, we estimate an uncertainty of 30\% for the total 
 cloud mass. 

For the star formation rate in the CNC, several recent determinations are available.
\citet{Povichetal2011} analyzed a \textit{Spitzer}-selected sample of YSOs
in the central 1.4~square-degrees of the Carina Nebula. From this they derived
a lower limit of $\dot{M} > 0.008\,M_\odot\,{\rm yr}^{-1}$ for the star formation
rate averaged over the past $\sim 2$~Myr. From an analysis of the optically visible
populations of stars in the Carina Nebula, they derived an average star formation rate 
over the past 5 Myr of $\dot{M} \approx 0.010 - 0.017\,M_\odot\,{\rm yr}^{-1}$. 
In our analysis of the YSO detected as point-like sources in our
\textit{Herschel} data of the entire CNC (i.e., including the Gum~31 region)
we estimated a star formation rate $0.017\,M_\odot\,{\rm yr}^{-1}$ \citep{Gaczkowskietal2013}, which 
is an average over the last few $10^5$~years. The error associated to the star formation rate 
is about 50\% \citep[details in ][]{Gaczkowskietal2013}. 
Taking the smaller area covered by the \textit{Spitzer} studies into account,
it can be shown \citep[see][for details]{Gaczkowskietal2013}  
that our \textit{Herschel}-based star formation rate  estimate
is in good agreement with the rates derived by \citet{Povichetal2011}.

\begin{figure}
\centering
\includegraphics[width=8.5cm]{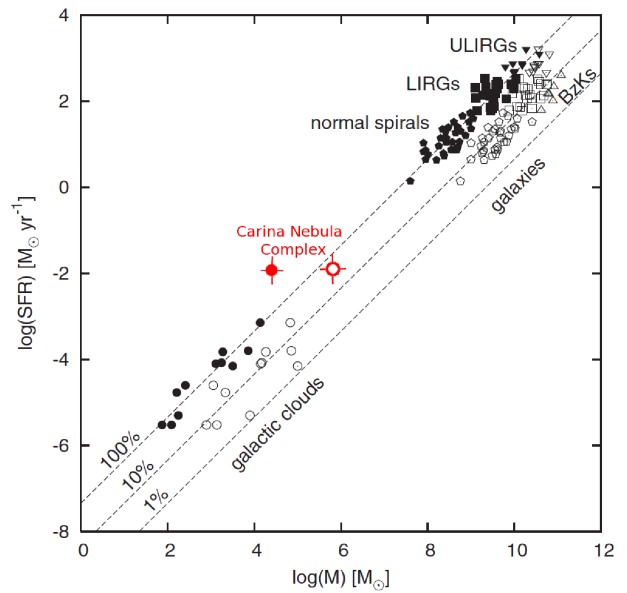}
   \caption{Plot adopted from Fig.2 in \citet{Ladaetal2012}. The red points represent the 
   position of the star formation rate and the total mass of the entire Carina Nebula 
   Complex. The error in the $\log\,(M)$ is 0.1 (i.e. $\sim$30\% of the total mass), and the error in 
   $\log\,(SFR)$ is 0.2 (i.e. $\sim$50\% of the star formation rate). The empty points represent the cloud 
   mass computed over an extinction threshold of A$_K>0.1$ and the filled point over A$_K>0.8$ 
   \citep[as in ][]{Ladaetal2012}.  three parallel
lines correspond to $f_{\rm DG} =$ 1.0, 0.1, and 0.01 (defined in Equation~\ref{sfr}) from left to right,
respectively.
   }
     \label{lada}
\end{figure}

In Figure~\ref{lada} we add our data for the CNC to the
star formation rate versus  cloud mass diagram from \citet{Ladaetal2012}.
Since the cloud mass as well as star formation rate of the CNC 
are larger than in all other Galactic regions shown in the original plot, the CNC 
adds an important new data point in the gap between the other Galactic regions
and the external galaxies.

Considering the total cloud mass above the $A_K>0.1$~mag threshold, 
the data point for the CNC appears to follow the general relation.
The cloud mass above the  $A_K>0.8$~mag threshold, however, appears to be 
considerably lower than the expectation according to the general relation.
This implies that the star formation rate per unit mass in dense clouds is
higher in the CNC compared to the other regions.
In other words, the
star formation process seems to be exceptionally efficient in the CNC.

Dividing the total mass in dense clouds of
$23\,000\,M_\odot$ by the star formation rate of $0.017\,M_\odot\,{\rm yr}^{-1}$ yields
 a characteristic timescale of 1.3~Myr.  If one would simply assume that the total mass of the
currently present dense clouds will be turned into stars at a temporally constant rate,
this process would last 1.3~Myr.
Considering the clearly established fact that star formation in the CNC is going on since
(at least) about 5~Myr (as witnessed by the large populations of young stars in the optically
visible populations), this would imply that the star formation process in the CNC
will soon come to its close and the observed dense clouds are the last remaining parts of the
original clouds.

However, we believe that it 
is probably not correct to think in terms of a fixed amount of dense clouds
that just turns into stars over time.
Rather, a large fraction of the dense clouds that are present today, may have just recently been
actively created by the feedback of the massive stars.

Several studies of the youngest stellar populations in the CNC came to the
conclusion that the currently ongoing star formation process is, to a large degree,
 spatially restricted to the edges and surfaces of irradiated clouds,
e.g.~the tips of the numerous pillars \citep{Smithetal2010, 
Ohlendorfetal2012a, Gaczkowskietal2013}.
 This suggests a causal connection between the star formation process and
local cloud compression. The ongoing feedback continues to compress (moderately dense) clouds
and thus can create new dense cloud structures that, once they get gravitationally
unstable, will collapse and form stars.

In this dynamic picture, where dense cloud structures are constantly created,
the ratio of dense cloud mass versus the star formation rate is naturally lower
than in the case of clouds that do not experience significant levels of
compression from massive star feedback.
This may explain the position of the CNC in this plot.
The CNC shows the largest deviation from the SFR versus cloud mass relation
than all other plotted Galactic star forming regions because the level of massive star
feedback is particularly high.

It is interesting to note some of the ``(ultra-) luminous infrared galaxies''
show deviations in the same sense, i.e.~SFRs that are higher than expected form the dense
cloud mass. This is consistent with the idea that a high level of feedback is increasing
the efficiency of star formation.

Since numerous galactic star forming regions have now been observed
by \textit{Herschel}, the analysis of these data will soon show
whether similarly large star forming complexes with comparable
levels of feedback do also show a similar behavior between
cloud mass and star formation rate. 

\begin{table*}
\caption{Cloud mass budget (in M$_\odot$) using different methods.}
\label{mass_methods}      
\centering               
\begin{tabular}{l l r rr}  
\hline\hline                
\noalign{\smallskip}
\multicolumn{2}{l}{Trace}     &  \multicolumn{2}{c}{Area}&Ref.\\
\noalign{\smallskip}
\hline                
\noalign{\smallskip}

 			&	&2.3$^\circ\times$2.3$^\circ$  &1.2$^\circ\times$1.2$^\circ$ &\\
\noalign{\smallskip}
\hline                
\noalign{\smallskip}
\multicolumn{2}{l}{Total NIR-Radio SED}   		                     &          900 000&&2\\
\noalign{\smallskip}
\hline                
\noalign{\smallskip}
\multicolumn{2}{l}{FIR} &&&\\ 
\noalign{\smallskip}
\hline                
\noalign{\smallskip}
		& SED - A$_{\rm V}>0.9$      &    609 765 &147 259&1\\
		& SED - A$_{\rm V}>1.0$       &         421 704 &128 220&1\\
		& SED - A$_{\rm V}>7.3$      &            22 903&  8 115 &1\\
\noalign{\smallskip}
\hline                
\noalign{\smallskip}
\multicolumn{2}{l}{Sub-mm/mm} &&&\\ 
\noalign{\smallskip}
\hline                
\noalign{\smallskip}
		& LABOCA                          &    &       60 000  &3\\
		& LABOCA (Clumps)        &    &        42 000&4\\
		& $^{12}$CO         	                     &   346 000 &     141 000&5\\
		& $^{13}$CO         	         	       &    132 000 &        63 000& 5\\
		& C$^{18}$O         		                &  57 800 &       22 000 &5\\
\noalign{\smallskip}
\hline                        
\noalign{\smallskip}
\end{tabular}
\tablebib{
(1) This work; 
(2) \citet[][]{Preibischetal2011c}; 
(3) \citet[][]{Preibischetal2012} (Paper I); 
(4) \citet[][]{Pekruhletal2013};  
(5) \citet[][]{Yonekuraetal2005}.
}
\end{table*}

\section{Conclusions and Summary}
\label{conclusions}

Our wide-field {\it Herschel} SPIRE and PACS maps allowed us to determine the temperatures,
surface densities, and the local strength of the FUV irradiation for all
cloud structures over the entire spatial extent ($\approx 125$~pc in diameter)
of the Carina Nebula complex at a spatial resolution of $\la 0.4$~pc for the first time.
The main results and conclusions of our study can be summarized as follows:

1. The global temperature structure of the clouds in the CNC is clearly dominated
by the radiative feedback from the numerous massive stars.

2. We present detailed temperature, column density, and FUV intensity maps
 of particularly interesting  small-scale cloud structures in the CNC.

3. The comparison of the {\it Herschel} FIR maps to a sub-mm map of the CNC shows
that the cold gas, traced by the sub-mm emission, is more spatially concentrated 
than the warmer gas, traced by {\it Herschel}.

4.  Considering the masses of individual clumps determined from their FIR dust emission,
we find that previous mass estimates that were
obtained from CO molecular line observations yielded systematically (by factors of $\sim 2-4$)
larger values. This difference probably results from uncertainties in the
absolute calibration of the different mass tracers and from the different biases 
(and beam sizes) of the observations.

5. We investigate the total cloud mass budget in the CNC at different column density thresholds
 and find that the different tracers (dust and CO) yield consistent total masses for
the moderately dense gas (as traced by $^{12}$CO) and the dense gas (as traced by $^{13}$CO and
C$^{18}$O).

6. The intensity of the FUV irradiation  at most cloud surfaces in the CNC
 is at least about 1000 times stronger than the galactic average value
(the so-called Habing field). At some locations, e.g.~at the surface of the
massive cloud to the west of the stellar cluster Tr~14, the FUV irradiation
is more than 10\,000 times the Habing field.

7. In a region north of the Carina Nebula, we discover a ``wave''-like pattern 
  in the column-density map. This pattern may be related to 
 a flow of hot gas streaming out of the northern part of the bipolar superbubble.

8. We compare the ratio between the total cloud mass and the star formation rate
in the CNC to the sample of Galactic and extragalactic star forming regions
originally presented by Lada et al.~(2012). This shows that
the CNC seems to form stars with a higher efficiency than other, 
typically lower-mass, Galactic star forming regions.
We suggest that this is a consequence of the particularly strong radiative feedback in the CNC.
The compression of the clouds actively and continuously creates dense clouds out of
lower density diffuse clouds. In this way, the total amount of dense clouds is
continuously replenished.

9. We present the first detailed FIR mapping and mass estimates for two cloud
complexes seen at the edges of our {\it Herschel} maps, which are molecular clouds
in the Galactic background at distances between 7 and 9~kpc.

\begin{acknowledgements}
The analysis of the {\it Herschel} data was funded by the German
Federal Ministry of Economics and Technology
in the framework of the  ``Verbundforschung Astronomie und Astrophysik''
through the DLR grant number 50~OR~1109. 
V.R. was partially supported by the {\it Bayerischen Gleichstellungsf{\"o}rderung} (BGF). 
Additional support came from funds from the Munich
Cluster of Excellence: ``Origin and Structure of the Universe''.
We acknowledge the HCSS / HSpot / HIPE, which are joint developments by the {\it Herschel} 
Science Ground Segment Consortium, consisting of ESA, the NASA {\it Herschel} Science Center, and the HIFI, 
PACS and SPIRE consortia. 
This publication uses data acquired with the Atacama Pathfinder
Experiment (APEX), which is a collaboration between
the Max-Planck-Institut f\"ur Radioastronomie, the European
Southern Observatory, and the Onsala Space Observatory. 
V.R. thanks A. Stutz for her help with the images convolution, H. Linz and M. Nielbock  for their advises and 
useful discussions. 
We thank C. Lada for the permission to reproduce Fig.10.
\end{acknowledgements}

\bibliographystyle{aa}
\bibliography{references}

\begin{thebibliography}{66}
\expandafter\ifx\csname natexlab\endcsname\relax\def\natexlab#1{#1}\fi

\bibitem[{{Adams} {et~al.}(2004){Adams}, {Hollenbach}, {Laughlin}, \&
  {Gorti}}]{Adamsetal2004}
{Adams}, F.~C., {Hollenbach}, D., {Laughlin}, G., \& {Gorti}, U. 2004, \apj,
  611, 360

\bibitem[{{Aniano} {et~al.}(2011){Aniano}, {Draine}, {Gordon}, \&
  {Sandstrom}}]{Anianoetal2011}
{Aniano}, G., {Draine}, B.~T., {Gordon}, K.~D., \& {Sandstrom}, K. 2011, \pasp,
  123, 1218

\bibitem[{{Bally} {et~al.}(2000){Bally}, {O'Dell}, \&
  {McCaughrean}}]{Ballyetal2000}
{Bally}, J., {O'Dell}, C.~R., \& {McCaughrean}, M.~J. 2000, \aj, 119, 2919

\bibitem[{{Bern{\'e}} {et~al.}(2010){Bern{\'e}}, {Marcelino}, \&
  {Cernicharo}}]{Berneetal2010}
{Bern{\'e}}, O., {Marcelino}, N., \& {Cernicharo}, J. 2010, \nat, 466, 947

\bibitem[{{Bronfman} {et~al.}(1996){Bronfman}, {Nyman}, \&
  {May}}]{Bronfmanetal1996}
{Bronfman}, L., {Nyman}, L.-A., \& {May}, J. 1996, \aaps, 115, 81

\bibitem[{{Brooks} {et~al.}(2003){Brooks}, {Cox}, {Schneider}, {Storey},
  {Poglitsch}, {Geis}, \& {Bronfman}}]{Brooksetal2003}
{Brooks}, K.~J., {Cox}, P., {Schneider}, N., {et~al.} 2003, \aap, 412, 751

\bibitem[{Burkert \& Hartmann(2012)}]{BurkertHartmann2012}
Burkert, A. \& Hartmann, L. 2012

\bibitem[{{Caswell} \& {Haynes}(1987)}]{CaswellHaynes1987}
{Caswell}, J.~L. \& {Haynes}, R.~F. 1987, \aap, 171, 261

\bibitem[{{Cersosimo} {et~al.}(2009){Cersosimo}, {Mader}, {Figueroa},
  {V{\'e}lez}, {Soto}, \& {Azc{\'a}rate}}]{Cersosimoetal2009}
{Cersosimo}, J.~C., {Mader}, S., {Figueroa}, N.~S., {et~al.} 2009, \apj, 699,
  469

\bibitem[{{Clarke}(2007)}]{Clarke2007}
{Clarke}, C.~J. 2007, \mnras, 376, 1350

\bibitem[{{Deharveng} {et~al.}(2005){Deharveng}, {Zavagno}, \&
  {Caplan}}]{Deharvengetal2005}
{Deharveng}, L., {Zavagno}, A., \& {Caplan}, J. 2005, \aap, 433, 565

\bibitem[{{Draine} \& {Lee}(1984)}]{DraineLee1984}
{Draine}, B.~T. \& {Lee}, H.~M. 1984, \apj, 285, 89

\bibitem[{{Fang} {et~al.}(2012){Fang}, {van Boekel}, {King}, {Henning},
  {Bouwman}, {Doi}, {Okamoto}, {Roccatagliata}, \&
  {Sicilia-Aguilar}}]{Fangetal2012a}
{Fang}, M., {van Boekel}, R., {King}, R.~R., {et~al.} 2012, \aap, 539, A119

\bibitem[{{Fontani} {et~al.}(2005){Fontani}, {Beltr{\'a}n}, {Brand},
  {Cesaroni}, {Testi}, {Molinari}, \& {Walmsley}}]{Fontanietal2005}
{Fontani}, F., {Beltr{\'a}n}, M.~T., {Brand}, J., {et~al.} 2005, \aap, 432, 921

\bibitem[{{Gaczkowski} {et~al.}(2013){Gaczkowski}, {Preibisch}, {Ratzka},
  {Roccatagliata}, {Ohlendorf}, \& {Zinnecker}}]{Gaczkowskietal2013}
{Gaczkowski}, B., {Preibisch}, T., {Ratzka}, T., {et~al.} 2013, A\&A, 549, A67

\bibitem[{{Goldsmith} {et~al.}(2008){Goldsmith}, {Heyer}, {Narayanan}, {Snell},
  {Li}, \& {Brunt}}]{Goldsmithetal2008}
{Goldsmith}, P.~F., {Heyer}, M., {Narayanan}, G., {et~al.} 2008, \apj, 680, 428

\bibitem[{{Griffin} {et~al.}(2010){Griffin}, {Abergel}, {Abreu}, {Ade},
  {Andr{\'e}}, {Augueres}, {Babbedge}, {Bae}, {Baillie}, {Baluteau}, {Barlow},
  {Bendo}, {Benielli}, {Bock}, {Bonhomme}, {Brisbin}, {Brockley-Blatt},
  {Caldwell}, {Cara}, {Castro-Rodriguez}, {Cerulli}, {Chanial}, {Chen},
  {Clark}, {Clements}, {Clerc}, {Coker}, {Communal}, {Conversi}, {Cox},
  {Crumb}, {Cunningham}, {Daly}, {Davis}, {de Antoni}, {Delderfield}, {Devin},
  {di Giorgio}, {Didschuns}, {Dohlen}, {Donati}, {Dowell}, {Dowell}, {Duband},
  {Dumaye}, {Emery}, {Ferlet}, {Ferrand}, {Fontignie}, {Fox}, {Franceschini},
  {Frerking}, {Fulton}, {Garcia}, {Gastaud}, {Gear}, {Glenn}, {Goizel},
  {Griffin}, {Grundy}, {Guest}, {Guillemet}, {Hargrave}, {Harwit}, {Hastings},
  {Hatziminaoglou}, {Herman}, {Hinde}, {Hristov}, {Huang}, {Imhof}, {Isaak},
  {Israelsson}, {Ivison}, {Jennings}, {Kiernan}, {King}, {Lange}, {Latter},
  {Laurent}, {Laurent}, {Leeks}, {Lellouch}, {Levenson}, {Li}, {Li},
  {Lilienthal}, {Lim}, {Liu}, {Lu}, {Madden}, {Mainetti}, {Marliani}, {McKay},
  {Mercier}, {Molinari}, {Morris}, {Moseley}, {Mulder}, {Mur}, {Naylor},
  {Nguyen}, {O'Halloran}, {Oliver}, {Olofsson}, {Olofsson}, {Orfei}, {Page},
  {Pain}, {Panuzzo}, {Papageorgiou}, {Parks}, {Parr-Burman}, {Pearce},
  {Pearson}, {P{\'e}rez-Fournon}, {Pinsard}, {Pisano}, {Podosek}, {Pohlen},
  {Polehampton}, {Pouliquen}, {Rigopoulou}, {Rizzo}, {Roseboom}, {Roussel},
  {Rowan-Robinson}, {Rownd}, {Saraceno}, {Sauvage}, {Savage}, {Savini},
  {Sawyer}, {Scharmberg}, {Schmitt}, {Schneider}, {Schulz}, {Schwartz},
  {Shafer}, {Shupe}, {Sibthorpe}, {Sidher}, {Smith}, {Smith}, {Smith},
  {Spencer}, {Stobie}, {Sudiwala}, {Sukhatme}, {Surace}, {Stevens}, {Swinyard},
  {Trichas}, {Tourette}, {Triou}, {Tseng}, {Tucker}, {Turner}, {Vaccari},
  {Valtchanov}, {Vigroux}, {Virique}, {Voellmer}, {Walker}, {Ward}, {Waskett},
  {Weilert}, {Wesson}, {White}, {Whitehouse}, {Wilson}, {Winter}, {Woodcraft},
  {Wright}, {Xu}, {Zavagno}, {Zemcov}, {Zhang}, \& {Zonca}}]{Griffinetal2010}
{Griffin}, M.~J., {Abergel}, A., {Abreu}, A., {et~al.} 2010, \aap, 518, L3

\bibitem[{{Griffith} \& {Wright}(1993)}]{GriffithWright1993}
{Griffith}, M.~R. \& {Wright}, A.~E. 1993, \aj, 105, 1666

\bibitem[{{Gritschneder} {et~al.}(2010){Gritschneder}, {Burkert}, {Naab}, \&
  {Walch}}]{Gritschnederetal2010}
{Gritschneder}, M., {Burkert}, A., {Naab}, T., \& {Walch}, S. 2010, \apj, 723,
  971

\bibitem[{{Gritschneder} {et~al.}(2009){Gritschneder}, {Naab}, {Walch},
  {Burkert}, \& {Heitsch}}]{Gritschnederetal2009}
{Gritschneder}, M., {Naab}, T., {Walch}, S., {Burkert}, A., \& {Heitsch}, F.
  2009, \apjl, 694, L26

\bibitem[{{Habing}(1968)}]{Habing1968}
{Habing}, H.~J. 1968, {Studies of physical conditions in H I regions.}

\bibitem[{{Hennemann} {et~al.}(2012){Hennemann}, {Motte}, {Schneider},
  {Didelon}, {Hill}, {Arzoumanian}, {Bontemps}, {Csengeri}, {Andr{\'e}},
  {Konyves}, {Louvet}, {Marston}, {Men'shchikov}, {Minier}, {Nguyen Luong},
  {Palmeirim}, {Peretto}, {Sauvage}, {Zavagno}, {Anderson}, {Bernard}, {Di
  Francesco}, {Elia}, {Li}, {Martin}, {Molinari}, {Pezzuto}, {Russeil}, {Rygl},
  {Schisano}, {Spinoglio}, {Sousbie}, {Ward-Thompson}, \&
  {White}}]{Hennemannetal2012}
{Hennemann}, M., {Motte}, F., {Schneider}, N., {et~al.} 2012, \aap, 543, L3

\bibitem[{{Hollenbach} {et~al.}(1991){Hollenbach}, {Takahashi}, \&
  {Tielens}}]{Hollenbachetal1991}
{Hollenbach}, D.~J., {Takahashi}, T., \& {Tielens}, A.~G.~G.~M. 1991, \apj,
  377, 192

\bibitem[{{Hollenbach} \& {Tielens}(1999)}]{HollenbachTielens1999}
{Hollenbach}, D.~J. \& {Tielens}, A.~G.~G.~M. 1999, Reviews of Modern Physics,
  71, 173

\bibitem[{{Kramer} {et~al.}(2008){Kramer}, {Cubick}, {R{\"o}llig}, {Sun},
  {Yonekura}, {Aravena}, {Bensch}, {Bertoldi}, {Bronfman}, {Fujishita},
  {Fukui}, {Graf}, {Hitschfeld}, {Honingh}, {Ito}, {Jakob}, {Jacobs}, {Klein},
  {Koo}, {May}, {Miller}, {Miyamoto}, {Mizuno}, {Onishi}, {Park}, {Pineda},
  {Rabanus}, {Sasago}, {Schieder}, {Simon}, {Stutzki}, {Volgenau}, \&
  {Yamamoto}}]{Krameretal2008}
{Kramer}, C., {Cubick}, M., {R{\"o}llig}, M., {et~al.} 2008, \aap, 477, 547

\bibitem[{{Lada} {et~al.}(2012){Lada}, {Forbrich}, {Lombardi}, \&
  {Alves}}]{Ladaetal2012}
{Lada}, C.~J., {Forbrich}, J., {Lombardi}, M., \& {Alves}, J.~F. 2012, \apj,
  745, 190

\bibitem[{{Minier} {et~al.}(2013){Minier}, {Tremblin}, {Hill}, {Motte},
  {Andr{\'e}}, {Lo}, {Schneider}, {Audit}, {White}, {Hennemann}, {Cunningham},
  {Deharveng}, {Didelon}, {Di Francesco}, {Elia}, {Giannini}, {Nguyen Luong},
  {Pezzuto}, {Rygl}, {Spinoglio}, {Ward-Thompson}, \&
  {Zavagno}}]{Minieretal2013}
{Minier}, V., {Tremblin}, P., {Hill}, T., {et~al.} 2013, \aap, 550, A50

\bibitem[{{Mottram} {et~al.}(2007){Mottram}, {Hoare}, {Lumsden}, {Oudmaijer},
  {Urquhart}, {Sheret}, {Clarke}, \& {Allsopp}}]{Mottrametal2007}
{Mottram}, J.~C., {Hoare}, M.~G., {Lumsden}, S.~L., {et~al.} 2007, \aap, 476,
  1019

\bibitem[{{O'dell} {et~al.}(1993){O'dell}, {Wen}, \& {Hu}}]{Odelletal1993}
{O'dell}, C.~R., {Wen}, Z., \& {Hu}, X. 1993, \apj, 410, 696

\bibitem[{{Ohlendorf} {et~al.}(2012){Ohlendorf}, {Preibisch}, {Gaczkowski},
  {Ratzka}, {Grellmann}, \& {McLeod}}]{Ohlendorfetal2012a}
{Ohlendorf}, H., {Preibisch}, T., {Gaczkowski}, B., {et~al.} 2012, \aap, 540,
  A81

\bibitem[{{Ohlendorf} {et~al.}(2013){Ohlendorf}, {Preibisch}, {Gaczkowski},
  {Ratzka}, {Ngoumou}, {Roccatagliata}, \& {Grellmann}}]{Ohlendorfetal2013}
{Ohlendorf}, H., {Preibisch}, T., {Gaczkowski}, B., {et~al.} 2013, \aap, 552,
  A14

\bibitem[{{Ossenkopf} \& {Henning}(1994)}]{OssenkopfHenning1994}
{Ossenkopf}, V. \& {Henning}, T. 1994, \aap, 291, 943

\bibitem[{{Ott}(2010)}]{Ott2010}
{Ott}, S. 2010, in Astronomical Society of the Pacific Conference Series, Vol.
  434, Astronomical Data Analysis Software and Systems XIX, ed. {Y.~Mizumoto,
  K.-I.~Morita, M.~Ohishi}, 139

\bibitem[{{Palmeirim} {et~al.}(2013){Palmeirim}, {Andr{\'e}}, {Kirk},
  {Ward-Thompson}, {Arzoumanian}, {K{\"o}nyves}, {Didelon}, {Schneider},
  {Benedettini}, {Bontemps}, {Di Francesco}, {Elia}, {Griffin}, {Hennemann},
  {Hill}, {Martin}, {Men'shchikov}, {Molinari}, {Motte}, {Nguyen Luong},
  {Nutter}, {Peretto}, {Pezzuto}, {Roy}, {Rygl}, {Spinoglio}, \&
  {White}}]{Palmeirimetal2013}
{Palmeirim}, P., {Andr{\'e}}, P., {Kirk}, J., {et~al.} 2013, \aap, 550, A38

\bibitem[{{Pekruhl} {et~al.}(2013){Pekruhl}, {Preibisch}, {Schuller}, \&
  {Menten}}]{Pekruhletal2013}
{Pekruhl}, S., {Preibisch}, T., {Schuller}, F., \& {Menten}, K. 2013, \aap,
  550, A29

\bibitem[{{Pilbratt} {et~al.}(2010){Pilbratt}, {Riedinger}, {Passvogel},
  {Crone}, {Doyle}, {Gageur}, {Heras}, {Jewell}, {Metcalfe}, {Ott}, \&
  {Schmidt}}]{Pilbrattetal2010}
{Pilbratt}, G.~L., {Riedinger}, J.~R., {Passvogel}, T., {et~al.} 2010, \aap,
  518, L1

\bibitem[{{Poglitsch} {et~al.}(2010){Poglitsch}, {Waelkens}, {Geis},
  {Feuchtgruber}, {Vandenbussche}, {Rodriguez}, {Krause}, {Renotte}, {van
  Hoof}, {Saraceno}, {Cepa}, {Kerschbaum}, {Agn{\`e}se}, {Ali}, {Altieri},
  {Andreani}, {Augueres}, {Balog}, {Barl}, {Bauer}, {Belbachir}, {Benedettini},
  {Billot}, {Boulade}, {Bischof}, {Blommaert}, {Callut}, {Cara}, {Cerulli},
  {Cesarsky}, {Contursi}, {Creten}, {De Meester}, {Doublier}, {Doumayrou},
  {Duband}, {Exter}, {Genzel}, {Gillis}, {Gr{\"o}zinger}, {Henning},
  {Herreros}, {Huygen}, {Inguscio}, {Jakob}, {Jamar}, {Jean}, {de Jong},
  {Katterloher}, {Kiss}, {Klaas}, {Lemke}, {Lutz}, {Madden}, {Marquet},
  {Martignac}, {Mazy}, {Merken}, {Montfort}, {Morbidelli}, {M{\"u}ller},
  {Nielbock}, {Okumura}, {Orfei}, {Ottensamer}, {Pezzuto}, {Popesso},
  {Putzeys}, {Regibo}, {Reveret}, {Royer}, {Sauvage}, {Schreiber}, {Stegmaier},
  {Schmitt}, {Schubert}, {Sturm}, {Thiel}, {Tofani}, {Vavrek}, {Wetzstein},
  {Wieprecht}, \& {Wiezorrek}}]{Poglitschetal2010}
{Poglitsch}, A., {Waelkens}, C., {Geis}, N., {et~al.} 2010, \aap, 518, L2

\bibitem[{{Povich} {et~al.}(2011){Povich}, {Smith}, {Majewski}, {Getman},
  {Townsley}, {Babler}, {Broos}, {Indebetouw}, {Meade}, {Robitaille},
  {Stassun}, {Whitney}, {Yonekura}, \& {Fukui}}]{Povichetal2011}
{Povich}, M.~S., {Smith}, N., {Majewski}, S.~R., {et~al.} 2011, \apjs, 194, 14

\bibitem[{{Preibisch} {et~al.}(2011{\natexlab{a}}){Preibisch}, {Hodgkin},
  {Irwin}, {Lewis}, {King}, {McCaughrean}, {Zinnecker}, {Townsley}, \&
  {Broos}}]{Preibischetal2011a}
{Preibisch}, T., {Hodgkin}, S., {Irwin}, M., {et~al.} 2011{\natexlab{a}},
  \apjs, 194, 10

\bibitem[{{Preibisch} {et~al.}(2011{\natexlab{b}}){Preibisch}, {Ratzka},
  {Gehring}, {Ohlendorf}, {Zinnecker}, {King}, {McCaughrean}, \&
  {Lewis}}]{Preibischetal2011b}
{Preibisch}, T., {Ratzka}, T., {Gehring}, T., {et~al.} 2011{\natexlab{b}},
  \aap, 530, A40

\bibitem[{{Preibisch} {et~al.}(2011{\natexlab{c}}){Preibisch}, {Ratzka},
  {Kuderna}, {Ohlendorf}, {King}, {Hodgkin}, {Irwin}, {Lewis}, {McCaughrean},
  \& {Zinnecker}}]{Preibischetal2011d}
{Preibisch}, T., {Ratzka}, T., {Kuderna}, B., {et~al.} 2011{\natexlab{c}},
  \aap, 530, A34

\bibitem[{{Preibisch} {et~al.}(2012){Preibisch}, {Roccatagliata}, {Gaczkowski},
  \& {Ratzka}}]{Preibischetal2012}
{Preibisch}, T., {Roccatagliata}, V., {Gaczkowski}, B., \& {Ratzka}, T. 2012,
  \aap, 541, A132 (Paper I)

\bibitem[{{Preibisch} {et~al.}(2011{\natexlab{d}}){Preibisch}, {Schuller},
  {Ohlendorf}, {Pekruhl}, {Menten}, \& {Zinnecker}}]{Preibischetal2011c}
{Preibisch}, T., {Schuller}, F., {Ohlendorf}, H., {et~al.} 2011{\natexlab{d}},
  \aap, 525, A92

\bibitem[{{Ricci} {et~al.}(2008){Ricci}, {Robberto}, \&
  {Soderblom}}]{Riccietal2008}
{Ricci}, L., {Robberto}, M., \& {Soderblom}, D.~R. 2008, \aj, 136, 2136

\bibitem[{{Richling} \& {Yorke}(2000)}]{RichlingYorke2000}
{Richling}, S. \& {Yorke}, H.~W. 2000, \apj, 539, 258

\bibitem[{Rivera-Ingraham {et~al.}(2013)Rivera-Ingraham, Martin, Polychroni,
  Motte, Schneider, Bontemps, Hennemann, Menshchikov, Luong, Andre,
  Arzoumanian, Bernard, Francesco, Elia, Fallscheer, Hill, Li, Minier, Pezzuto,
  Roy, Rygl, Sadavoy, Spinoglio, White, \& Wilson}]{Rivera-Ingrahametal2013}
Rivera-Ingraham, A., Martin, P.~G., Polychroni, D., {et~al.} 2013, \apj, in
  press

\bibitem[{{Roccatagliata} {et~al.}(2011){Roccatagliata}, {Bouwman}, {Henning},
  {Gennaro}, {Feigelson}, {Kim}, {Sicilia-Aguilar}, \&
  {Lawson}}]{Roccatagliataetal2011}
{Roccatagliata}, V., {Bouwman}, J., {Henning}, T., {et~al.} 2011, \apj, 733,
  113

\bibitem[{{Roussel}(2012)}]{Roussel2012}
{Roussel}, H. 2012, ArXiv 1205.2576

\bibitem[{{Russeil}(2003)}]{Russeil2003}
{Russeil}, D. 2003, \aap, 397, 133

\bibitem[{{Russeil} \& {Castets}(2004)}]{RusseilCastets2004}
{Russeil}, D. \& {Castets}, A. 2004, \aap, 417, 107

\bibitem[{{Schneider} \& {Brooks}(2004)}]{SchneiderBrooks2004}
{Schneider}, N. \& {Brooks}, K. 2004, \pasa, 21, 290

\bibitem[{{Schneider} {et~al.}(2010){Schneider}, {Motte}, {Bontemps},
  {Hennemann}, {di Francesco}, {Andr{\'e}}, {Zavagno}, {Csengeri},
  {Men'shchikov}, {Abergel}, {Baluteau}, {Bernard}, {Cox}, {Didelon}, {di
  Giorgio}, {Gastaud}, {Griffin}, {Hargrave}, {Hill}, {Huang}, {Kirk},
  {K{\"o}nyves}, {Leeks}, {Li}, {Marston}, {Martin}, {Minier}, {Molinari},
  {Olofsson}, {Panuzzo}, {Persi}, {Pezzuto}, {Roussel}, {Russeil}, {Sadavoy},
  {Saraceno}, {Sauvage}, {Sibthorpe}, {Spinoglio}, {Testi}, {Teyssier},
  {Vavrek}, {Ward-Thompson}, {White}, {Wilson}, \&
  {Woodcraft}}]{Schneideretal2010}
{Schneider}, N., {Motte}, F., {Bontemps}, S., {et~al.} 2010, \aap, 518, L83

\bibitem[{{Schneider} {et~al.}(1998){Schneider}, {Stutzki}, {Winnewisser},
  {Poglitsch}, \& {Madden}}]{Schneideretal1998}
{Schneider}, N., {Stutzki}, J., {Winnewisser}, G., {Poglitsch}, A., \&
  {Madden}, S. 1998, \aap, 338, 262

\bibitem[{{Smith}(2006)}]{Smith2006}
{Smith}, N. 2006, \mnras, 367, 763

\bibitem[{{Smith} \& {Brooks}(2007)}]{SmithBrooks2007}
{Smith}, N. \& {Brooks}, K.~J. 2007, \mnras, 379, 1279

\bibitem[{{Smith} \& {Brooks}(2008)}]{SmithBrooks2008}
{Smith}, N. \& {Brooks}, K.~J. 2008, {in Handbook of Star Forming Regions,
  Volume II: The Southern Sky}, ed. {Reipurth, B.}, 138

\bibitem[{{Smith} \& {Conti}(2008)}]{SmithConti2008}
{Smith}, N. \& {Conti}, P.~S. 2008, \apj, 679, 1467

\bibitem[{{Smith} {et~al.}(2000){Smith}, {Egan}, {Carey}, {Price}, {Morse}, \&
  {Price}}]{Smithetal2000}
{Smith}, N., {Egan}, M.~P., {Carey}, S., {et~al.} 2000, \apjl, 532, L145

\bibitem[{{Smith} {et~al.}(2010){Smith}, {Povich}, {Whitney}, {Churchwell},
  {Babler}, {Meade}, {Bally}, {Gehrz}, {Robitaille}, \&
  {Stassun}}]{Smithetal2010}
{Smith}, N., {Povich}, M.~S., {Whitney}, B.~A., {et~al.} 2010, \mnras, 406, 952

\bibitem[{{Smith} {et~al.}(2005){Smith}, {Stassun}, \& {Bally}}]{Smithetal2005}
{Smith}, N., {Stassun}, K.~G., \& {Bally}, J. 2005, \aj, 129, 888

\bibitem[{{Stutz} {et~al.}(2010){Stutz}, {Launhardt}, {Linz}, {Krause},
  {Henning}, {Kainulainen}, {Nielbock}, {Steinacker}, \&
  {Andr{\'e}}}]{Stutzetal2010}
{Stutz}, A., {Launhardt}, R., {Linz}, H., {et~al.} 2010, \aap, 518, L87

\bibitem[{{Townsley} {et~al.}(2011{\natexlab{a}}){Townsley}, {Broos}, {Chu},
  {Gagn{\'e}}, {Garmire}, {Gruendl}, {Hamaguchi}, {Mac Low}, {Montmerle},
  {Naz{\'e}}, {Oey}, {Park}, {Petre}, \& {Pittard}}]{Townsleyetal2011a}
{Townsley}, L.~K., {Broos}, P.~S., {Chu}, Y.-H., {et~al.} 2011{\natexlab{a}},
  \apjs, 194, 15

\bibitem[{{Townsley} {et~al.}(2011{\natexlab{b}}){Townsley}, {Broos},
  {Corcoran}, {Feigelson}, {Gagn{\'e}}, {Montmerle}, {Oey}, {Smith}, {Garmire},
  {Getman}, {Povich}, {Remage Evans}, {Naz{\'e}}, {Parkin}, {Preibisch},
  {Wang}, {Wolk}, {Chu}, {Cohen}, {Gruendl}, {Hamaguchi}, {King}, {Mac Low},
  {McCaughrean}, {Moffat}, {Oskinova}, {Pittard}, {Stassun}, {ud-Doula},
  {Walborn}, {Waldron}, {Churchwell}, {Nichols}, {Owocki}, \&
  {Schulz}}]{Townsleyetal2011}
{Townsley}, L.~K., {Broos}, P.~S., {Corcoran}, M.~F., {et~al.}
  2011{\natexlab{b}}, \apjs, 194, 1

\bibitem[{{Tremblin} {et~al.}(2012){Tremblin}, {Audit}, {Minier}, {Schmidt}, \&
  {Schneider}}]{Tremblinetal2012}
{Tremblin}, P., {Audit}, E., {Minier}, V., {Schmidt}, W., \& {Schneider}, N.
  2012, \aap, 546, A33

\bibitem[{{Vall{\'e}e}(2008)}]{Valleeetal2009}
{Vall{\'e}e}, J.~P. 2008, \aj, 135, 1301

\bibitem[{{Yonekura} {et~al.}(2005){Yonekura}, {Asayama}, {Kimura}, {Ogawa},
  {Kanai}, {Yamaguchi}, {Barnes}, \& {Fukui}}]{Yonekuraetal2005}
{Yonekura}, Y., {Asayama}, S., {Kimura}, K., {et~al.} 2005, \apj, 634, 476

\end{thebibliography}

\begin{appendix}

\section{Cloud complexes at the edges of the \textit{Herschel} maps}
Besides the clouds associated to the Carina Nebula and the
Gum~31 region, 
our \textit{Herschel} maps also show several cloud 
structures at the periphery of the field of view.
The two most prominent peripheral cloud complexes are seen near the
south-eastern edge and the western edge of our map.
Since no detailed FIR observations of these clouds seem to be available so far
in the literature, we briefly describe the FIR morphology of 
these cloud complexes as revealed by our \textit{Herschel} maps and discuss
identifications with observations at other wavelengths.


\subsection{The G$289.0-0.3$ cloud complex}  

A cloud complex extending over about $35' \times 35'$ is seen 
around the position
$(\alpha_{\rm J2000},\, \delta_{\rm J2000})\, =\, (10^{\rm h}\,56^{\rm m}\,30^{\rm s},\, -60\degr\,06'\,00'')$ ,
near the south-eastern edge of our SPIRE maps.
Optical images of this region show numerous field stars, but no prominent 
cloud features.
The cloud complex we see in the \textit{Herschel} maps is related to the 
radio-detected molecular cloud G289.0-0.3 described by \citet{RusseilCastets2004}.
With a radial velocity of $v_{\rm rad} = + 27.5$~km/s, this cloud 
complex seems to be clearly unrelated to the Carina Nebula cloud complex
(which has $v_{\rm rad} \approx - 26$~km/s).

\begin{figure*}
\centering
\includegraphics[width=9cm]{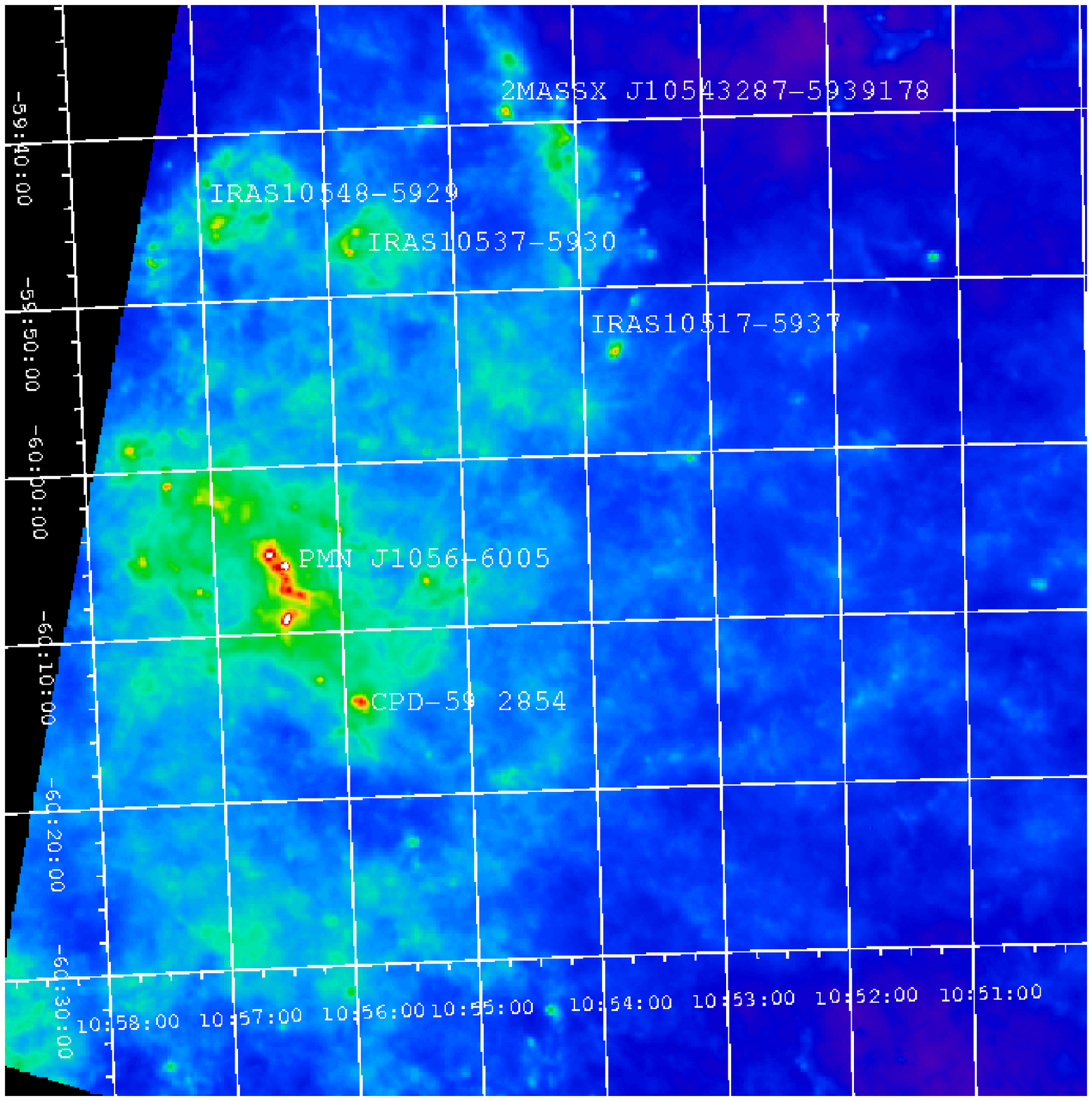}
\includegraphics[width=9cm]{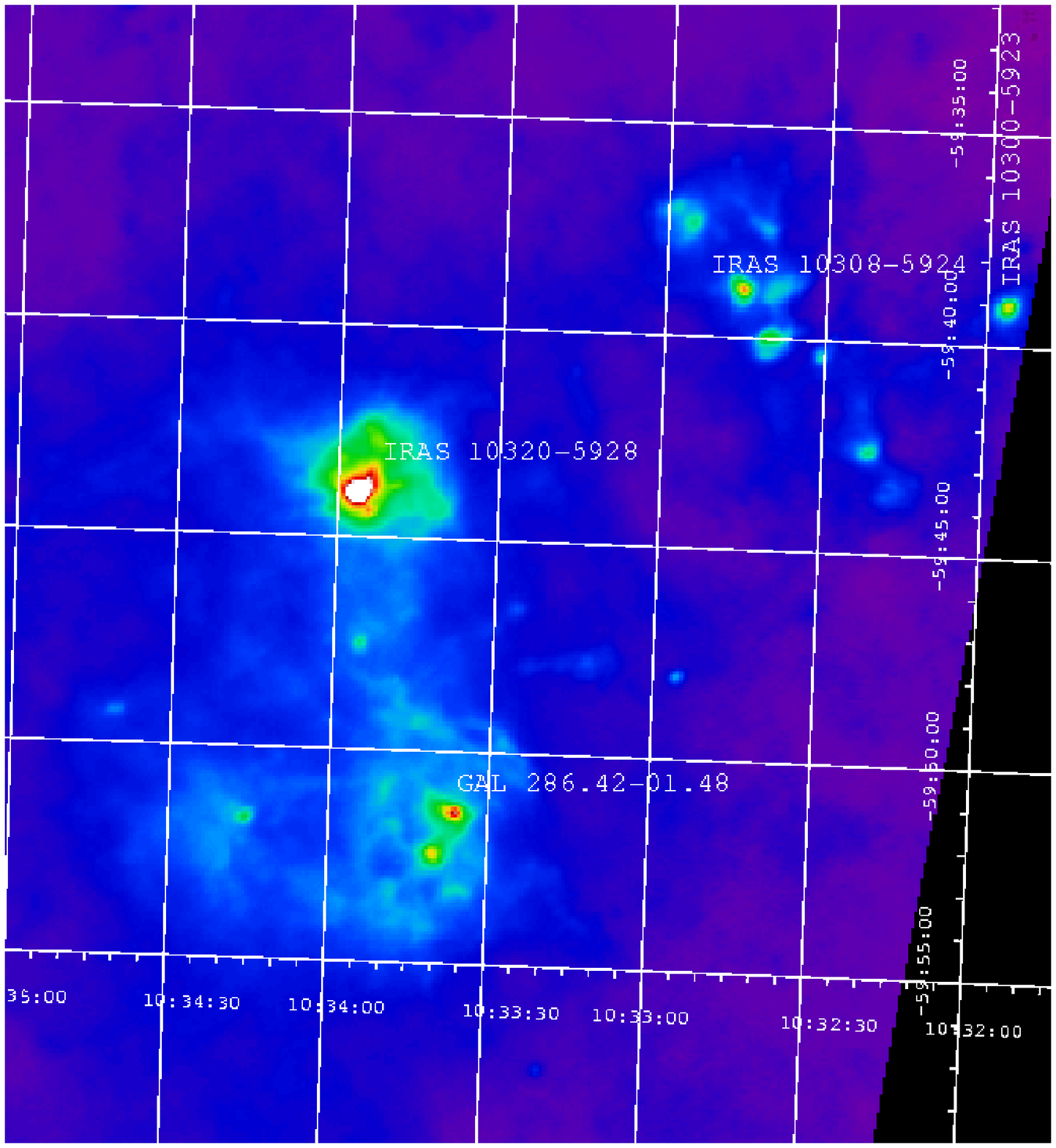}
  \caption{{\it Right:}  \textit{Herschel}\,/\,PACS $70\,\mu$m map of the G$286.4-1.3$ cloud complex; 
  	{\it Left:}  \textit{Herschel}\,/\,SPIRE $250\,\mu$m map of the G$289.0-0.3$ cloud complex. 
	}
      \label{new_clouds}
\end{figure*}

Besides the diffuse cloud emission, our \textit{Herschel} maps
revealed 52 individual point-like sources, the
positions and fluxes of which are given in \cite{Gaczkowskietal2013}.
Several of these can be identified with IRAS or MSX point-sources, some of which are
marked in Fig.~\ref{new_clouds}.
The brightest compact \textit{Herschel} source in this area can be
identified with the
giant H\,II region PMN~J1056-6005, which  was
detected in the Parkes-MIT-NRAO 4850 MHz survey
\citep{GriffithWright1993}, and seems to be identical to the H\,II region
GAL 289.06-00.36 listed in \citet{CaswellHaynes1987}.
\citet{Cersosimoetal2009} detected emission in the H166$\alpha$ 
radio recombination line from this region and determined a distance
of $D = 7.1 \pm 0.3$ kpc.
We note that this distance value is consistent with the prediction of the velocimetric model
of the Galaxy by \cite{Valleeetal2009} for the galactic longitude and the
measured radial velocity of this molecular cloud complex.

At the southern tip of the cloud complex, the
Be~star CPD--59~2854 (= IRAS~10538-5958) is seen as a bright compact
far-infrared source.
In the northern part of this complex, 
one of the bright  \textit{Herschel} point-like
sources is the extended 2MASS source
 2MASX J10543287-5939178, which is listed as a Galaxy in the
SIMBAD catalog, but may in fact be
an ultra-compact H\,II region  \citep{Bronfmanetal1996}.

Since this cloud complex is located near the edge of our {\it Herschel} maps,
and a considerable fraction is only covered by the SPIRE maps, but not
by PACS, we cannot provide a complete characterization of this cloud
complex. In our temperature map, the dust temperatures in these clouds
ranges from about 18~K to about 25~K.
Integrating our column density map (see
Section\ref{an}) over a $16' \times 19'$ area, covering that part
of the cloud complex that is observed by PACS \textit{and} SPIRE, 
we obtain a total mass of  $65\,000\,M_{\sun}$.
Due to the incomplete spatial coverage, this is clearly a lower limit to the 
total cloud mass of this complex.

This rather high cloud mass and the remarkably high number of detected point-like
\textit{Herschel} sources\footnote{We note that due to the larger distance, 
the \textit{Herschel} point-source detection limit 
corresponds to about 10 times larger FIR fluxes than for the young stellar objects 
in the Carina Nebula; this suggests that most of the point-like FIR sources 
detected in this cloud complex must be
relative massive ($\ga 10\,M_\odot$) protostellar objects (or, alternatively,
very compact and rich clusters of young stellar objects)}
clearly show that this cloud complex is a region of active massive star 
formation that is worth to be studied in more detail.

\subsection{The G$286.4-1.3$ cloud complex}  

Near the western edge of our PACS maps, a cloud complex
extending over about $16' \times 19'$ is seen around the position 
$(\alpha_{\rm J2000},\, \delta_{\rm J2000})\, =\, ( 10^{\rm h}\,34^{\rm m}\,00^{\rm s}, \,-59\degr\,47'\,00'')$. 
Optical images of this region 
show several dark clouds.

The cloud complex seems to be related to the
radio-detected molecular cloud G286.4-1.3 discussed by \citet{RusseilCastets2004}.
The radial velocity of $v_{\rm rad} = + 39.9$~km/s is clearly
distinct from that of the Carina Nebula cloud complex
($v_{\rm rad} \approx - 26$~km/s). 
We note that no CO emission from these clouds was detected in the 
survey of \citet{Yonekuraetal2005}.

Besides the diffuse cloud emission, our \textit{Herschel} maps
revealed 15 individual point-like sources in these clouds, the
positions and fluxes of which are given in \cite{Gaczkowskietal2013}.
The brightest emission peak in this complex 
can be identified with the
ultra-compact H\,II region IRAS~10320-5928
\citep{Bronfmanetal1996}.

The southern part of the cloud complex contains the 
FIR source IRAS~10317-5936, which is identified with the 
H\,II~region GAL 286.42-01.48 and  
listed as part of the star-forming complex 286.4-01.4 in \citet{Russeil2003}.
It contains the two candidate massive young stellar objects
286.3938-01.3514  1 and 2 \citep{Mottrametal2007}.
\citet{Fontanietal2005} 
determined a distance of $D = 8.88$~kpc to this cloud.
We note that this distance value is consistent with the prediction of the velocimetric model 
of the Galaxy by \cite{Valleeetal2009} for the galactic longitude and the
measured radial velocity of this molecular cloud complex.

In the north-western part of this complex,
another accumulation of compact clouds is found, two of which
can be identified with IRAS sources.

The main part, but not the full extent of this cloud complex is covered
by our PACS \textit{and} SPIRE maps. 
The dust temperatures range from about 20~K to $\ga 35$~K.
Integrating our column density map  yields a total
mass of $\sim 93\,000\,M_\odot$ within a $16' \times 19'$ area, if we assume
the above mentioned distance of 8.88~kpc. This is a strict lower limit, since
our data cover only some part of the cloud complex.

This rather high cloud mass and the presence of numerous bright point-like
\textit{Herschel} sources (with fluxes up to $\approx530$~Jy, corresponding to
FIR luminosities of up to $\sim 25\,000\,L_\odot$) shows that this complex is a site of
active massive star formation and certainly worth to be studied in more detail.

\end{appendix}

\end{document}